\documentclass[a4paper,twocolumn,11pt]{quantum/quantumarticle}

\pdfoutput=1
\usepackage{tikz}
\usepackage{geometry}
\usepackage{tikz}
\usepackage{tikz/tikzit}
\tikzstyle{gate}=[shape=rectangle, text height=1.5ex, text depth=0.25ex, yshift=0.5mm, fill=white, draw=black, minimum height=5mm, yshift=-0.5mm, minimum width=5mm, font={\normalsize}, tikzit category=circuit]
\tikzstyle{big gate}=[shape=rectangle, text height=1.5ex, text depth=0.25ex, yshift=0.5mm, fill=white, draw=black, minimum height=10mm, yshift=-0.5mm, minimum width=5mm, font={\normalsize}, tikzit category=circuit]
\tikzstyle{Z dot}=[inner sep=0mm, minimum size=2mm, shape=circle, draw=black, fill=spider green, tikzit category=zx]
\tikzstyle{Z phase dot}=[minimum size=5mm, font={\footnotesize\boldmath}, shape=rectangle, rounded corners=2mm, inner sep=0.2mm, outer sep=-2mm, scale=0.8, tikzit shape=circle, draw=black, fill=spider green, tikzit draw=blue, tikzit category=zx]
\tikzstyle{X dot}=[Z dot, shape=circle, draw=black, fill={rgb,255: red,255; green,136; blue,136}, tikzit category=zx]
\tikzstyle{X phase dot}=[Z phase dot, tikzit shape=circle, tikzit draw=blue, fill={rgb,255: red,255; green,136; blue,136}, font={\footnotesize\boldmath}, tikzit category=zx]
\tikzstyle{hadamard}=[fill=yellow, draw=black, shape=rectangle, inner sep=0.6mm, minimum height=1.5mm, minimum width=1.5mm, tikzit category=zx]
\tikzstyle{paulibox}=[fill={rgb,255: red,221; green,221; blue,255}, draw=black, shape=rectangle, inner sep=0.6mm, minimum height=5mm, minimum width=5mm, font={\footnotesize}, text height=1.5ex, text depth=0.25ex, tikzit category=zx]
\tikzstyle{vertex}=[inner sep=0mm, minimum size=1mm, shape=circle, draw=black, fill=black, tikzit category=misc]
\tikzstyle{vertex set}=[inner sep=0mm, minimum size=1mm, shape=circle, draw=black, fill=white, font={\footnotesize\boldmath}, tikzit category=misc]
\tikzstyle{small black dot}=[fill=black, draw=black, shape=circle, inner sep=0pt, minimum width=1.2mm, tikzit category=circuit]
\tikzstyle{cnot ctrl}=[fill=black, draw=black, shape=circle, inner sep=0pt, minimum width=1.2mm, tikzit category=circuit]
\tikzstyle{cnot targ}=[fill=white, draw=white, shape=circle, tikzit category=circuit, label={center:$\oplus$}, inner sep=0pt, minimum width=2.1mm, tikzit fill={rgb,255: red,102; green,204; blue,255}, tikzit draw=black]
\tikzstyle{ket}=[fill=white, draw=black, shape=regular polygon, regular polygon sides=3, regular polygon rotate=-30, scale=0.7, inner sep=1pt, tikzit category=circuit, tikzit shape=rectangle, tikzit fill=green]
\tikzstyle{bra}=[fill=white, draw=black, shape=regular polygon, regular polygon sides=3, regular polygon rotate=30, scale=0.7, inner sep=1pt, tikzit category=circuit, tikzit shape=rectangle, tikzit fill=red]
\tikzstyle{scalar}=[shape=rectangle, text height=1.5ex, text depth=0.25ex, yshift=0.5mm, fill=white, draw=black, minimum height=5mm, yshift=-0.5mm, minimum width=5mm, font={\normalsize}]
\tikzstyle{clabel}=[fill=white, draw=none, shape=rectangle, tikzit fill={rgb,255: red,56; green,255; blue,242}, font={\footnotesize}, inner sep=1pt, tikzit category=labels]
\tikzstyle{empty diagram}=[draw={gray!40!white}, dashed, shape=rectangle, minimum width=1cm, minimum height=1cm, tikzit category=misc]

\tikzstyle{hadamard edge}=[-, dashed, dash pattern=on 2pt off 0.5pt, thick, draw=luxembourg blue]
\tikzstyle{box edge}=[-, dashed, dash pattern=on 2pt off 0.5pt, thick, draw={rgb,255: red,203; green,192; blue,225}]
\tikzstyle{brace edge}=[-, tikzit draw=blue, decorate, decoration={brace,amplitude=1mm,raise=-1mm}]
\tikzstyle{diredge}=[->]
\tikzstyle{double edge}=[-, double, shorten <=-1mm, shorten >=-1mm, double distance=2pt]
\tikzstyle{gray edge}=[-, {gray!70!white}, thick]
\tikzstyle{pointer edge}=[->, very thick, gray]
\tikzstyle{boldedge}=[-, line width=1.6pt, shorten <=-0.17mm, shorten >=-0.17mm]

\definecolor{luxembourg blue}{HTML}{00a4dd}                                                                                                                          
\definecolor{luxembourg red}{HTML}{e31b1d}
\definecolor{spider green}{RGB}{221,255,221}
\definecolor{spider red}{RGB}{255,136,136}
\definecolor{luxembourg green}{rgb}{0.000000,0.501961,0.000000}
\definecolor{luxembourg orange}{rgb}{1.000000,0.647059,0.000000}%

\usetikzlibrary{calc}
\usetikzlibrary{intersections}
\usetikzlibrary{shapes,arrows,intersections,patterns}
\usetikzlibrary{calc, arrows.meta, intersections, patterns,
positioning, shapes.misc, fadings, through,decorations.pathreplacing}

\usetikzlibrary{matrix,fit,calc,trees,positioning,arrows,chains,shapes.geometric,shapes,zx-calculus}
\usepackage[usenames,svgnames,dvipsnames,table,x11names]{xcolor}
\DeclareGraphicsExtensions{.pdf,.jpeg,.png}
\usepackage{colortbl}
\usepackage{array}
\usepackage[numbers]{natbib}
\usepackage{booktabs}
\usepackage{blochsphere}
\usepackage{colortbl}
\usepackage{subcaption}
\usepackage{mathtools}
\usepackage{wrapfig}
\usepackage{amssymb}
\usepackage{hyperref}
\usepackage[utf8]{inputenc}
\usepackage[english]{babel}
\usepackage[T1]{fontenc}
\usepackage{pgf}
\usepackage{pgfplots}
\usepackage{nicematrix}
\usepackage{amsmath}
\usepackage{hyperref}
\usepackage{array}
\usepackage{tabularx}

\DeclarePairedDelimiter\bra{\langle}{\rvert}
\DeclarePairedDelimiter\ket{\lvert}{\rangle}
\DeclarePairedDelimiterX\braket[2]{\langle}{\rangle}{#1\,\delimsize\vert\,\mathopen{}#2}
\DeclarePairedDelimiter\abs{\lvert}{\rvert}%
\DeclarePairedDelimiter\norm{\lVert}{\rVert}%

\newcommand{\yestaxonomy}{%
  \tikz[scale=0.23,baseline=(current bounding box.south)] {
    \draw[line width=1.0,line cap=round] (0.25,0) to [bend left=10] (1,1);
    \draw[line width=0.9,line cap=round] (0,0.35) to [bend right=1] (0.23,0);
}}
\newcommand\identity{1\kern-0.25em\text{l}}

\newcommand{\mat}[1]{\mathit{\mathbf{#1}}}

\makeatletter
\let\oldabs\abs
\def\abs{\@ifstar{\oldabs}{\oldabs*}}
\let\oldnorm\norm
\def\norm{\@ifstar{\oldnorm}{\oldnorm*}}
\makeatother

\newcolumntype{C}[1]{>{\centering\arraybackslash}m{#1}}

\usepackage{amsthm}
\newtheorem{definition}{Definition}[section]

\newcommand{\blochstate}[1]{
  \begin{blochsphere}[radius=0.875 cm,tilt=15,rotation=-20,
    color=white, opacity = 0.1, ball=circle, ]
    \drawBallGrid[style={opacity=0.2}]{30}{30}

    \labelLatLon{up}{90}{0};
    \labelLatLon{down}{-90}{90};
    \labelLatLon{left}{-180}{0};
    \labelLatLon{right}{0}{0};
    \labelLatLon{front}{0}{90};
    \labelLatLon{behind}{0}{-90};
    \labelLatLon{state}{0}{90};

    \draw[-latex] (0,0) -- (up) node[above,inner sep=.5mm] at (up)
    {\footnotesize $\ket{0}$};
    \draw[-latex] (0,0) -- (down) node[below,inner sep=.5mm] at
    (down) {\footnotesize $\ket{1}$};

    \draw[-latex] (0,0) -- (left) node[left,inner sep=.5mm] at (left)
    {\footnotesize $\ket{-i}$};
    \draw[-latex] (0,0) -- (right) node[right,inner sep=.5mm] at
    (right) {\footnotesize $\ket{+i}$};

    \draw[-latex] (0,0) -- (front) node[below,inner sep=.5mm] at
    (front) {\footnotesize $\ket{+}$};
    \draw[-latex] (0,0) -- (behind) node[below,inner sep=.5mm] at
    (behind) {\footnotesize $\ket{-}$};

    \draw[red,ultra thick, -latex] (0,0) -- (#1) node[red,
    right,inner sep=2.5mm] at (#1) {$\mathbf{\ket{\Psi}}$};

  \end{blochsphere}
}

\usepackage{pifont}
\newcommand{\no}{%
\tikz[scale=0.23,baseline=(current bounding box.center)] {
    \draw[line width=1.0,line cap=round] (0,0) to [bend left=6] (1,1);
    \draw[line width=1.0,line cap=round] (0.2,0.95) to [bend right=3] (0.8,0.05);
    \draw[line width=0.7] (-0.2,-0.2) rectangle (1.2,1.2); 
}}
\newcommand{\yes}{%
\tikz[scale=0.23,baseline=(current bounding box.center)] {
    \draw[line width=1.0,line cap=round] (0.25,0) to [bend left=10] (1,1);
    \draw[line width=0.9,line cap=round] (0,0.35) to [bend right=1] (0.23,0);
    \draw[line width=0.7] (-0.2,-0.2) rectangle (1.2,1.2); 
}}

\usepackage{qcircuit}
\newcommand{\qcid}{\Qcircuit @C=.5em @R=.75em {& \gate{I} & \qw}}
\newcommand{\qcx}{\Qcircuit @C=.5em @R=.75em {& \gate{X} & \qw}}
\newcommand{\qcy}{\Qcircuit @C=.5em @R=.75em {& \gate{Y} & \qw}}
\newcommand{\qcz}{\Qcircuit @C=.5em @R=.75em {& \gate{Z} & \qw}}
\newcommand{\qcs}{\Qcircuit @C=.5em @R=.75em {& \gate{S} & \qw}}
\newcommand{\qcsd}{\Qcircuit @C=.5em @R=.75em {& \gate{S^\dagger} & \qw}}

\newcommand{\qch}{\Qcircuit @C=.5em @R=.75em {& \gate{H} & \qw}}
\newcommand{\qct}{\Qcircuit @C=.5em @R=.75em {& \gate{T} & \qw}}
\newcommand{\qctd}{\Qcircuit @C=.5em @R=.75em {& \gate{T^\dagger} & \qw}}
\newcommand{\qccx}{\Qcircuit @C=.5em @R=.75em {& \ctrl{1} & \qw \\ &
\targ & \qw}}

\newcommand{\qccz}{\Qcircuit @C=.5em @R=.75em {& \ctrl{1}   & \qw \\
& \gate{Z} \qw & \qw}}

\AtBeginDocument{%
  }

\begin{document}

\title{A Review on Quantum Circuit Optimization using ZX-Calculus}

\author{Tobias~Fischbach}
\email{tobias.fischbach@uni.lu}
\orcid{0009-0001-2535-2577}
\affiliation{Department of Computer Science, University of
Luxembourg, Esch-sur-Alzette, Luxembourg}
\author{Pierre~Talbot}
\orcid{0000-0001-9202-4541}
\affiliation{Department of Computer Science, University of
Luxembourg, Esch-sur-Alzette, Luxembourg}
\author{Pascal~Bouvry}
\orcid{0000-0001-9338-2834}
\affiliation{Department of Computer Science, University of
Luxembourg, Esch-sur-Alzette, Luxembourg}
%\authorrunning{T. Fischbach et al.}

%\renewcommand{\shortauthors}{Fischbach et al.}

\begin{abstract}
  Quantum computing promises significant speed-ups for certain
  algorithms but the practical use of current noisy intermediate-scale
  quantum (NISQ)-era computers remains limited by resources
  constraints (e.g., noise,
  qubits, gates, and circuit depth).
  Quantum circuit optimization is a key mitigation strategy.
  In this context, ZX-calculus has emerged as an
  alternative framework that allows for semantics-preserving quantum
  circuit optimization.

  We review ZX-based
  optimization of quantum circuits, categorizing them by optimization
  techniques, target metrics and intended quantum computing architecture.
  In addition, we outline critical challenges and future research
  directions, such as multi-objective optimization, scalable algorithms,
  and enhanced circuit extraction methods.

  %Given the interdisciplinary nature of ZX-based optimization, this
  %survey addresses researchers from different scientific communities.
  %The primary target audiences are researchers experienced with:
  %(i) quantum circuit optimization but
  %unfamiliar with ZX-calculus, (ii) combinatorial
  %optimization but with limited exposure to quantum computing,
  %and (iii) ZX-based methods who desire a comprehensive overview of
  %current optimization approaches.

  This survey is valuable for researchers in both combinatorial
  optimization and quantum computing. For researchers in
  combinatorial optimization, we provide the background to understand
  a new challenging combinatorial problem: ZX-based quantum circuit
  optimization. For researchers in quantum computing, we classify and
  explain existing circuit optimization techniques.

\end{abstract}

\maketitle

\section{Introduction}\label{sec:introduction}

\textit{Quantum computing} belongs to the field of quantum information
theory and uses quantum mechanical effects to process information.
As a consequence, new applications emerge that are intractable for classical
computers~\cite{gill2022surveyqc} (Section~\ref{sec:quantum_computing}).
Although quantum computing promises advances in drug discovery,
material science, and climate modeling, its advantages are not
universal across applications~\cite{hassijaForthcomingApplicationsQuantum2020}.
But as the quantum computing paradigm differs
fundamentally from classical computing, quantum algorithms need to
be carefully designed to exceed the performance of classical
computers~\cite{preskill2018nisq}.

A \textit{quantum
circuit}~\cite{barencoElementaryGatesQuantum1995a} is the standard
model to express
quantum algorithms.
Its basic building blocks are qubits and quantum gates.
The \textit{qubit} is the elementary unit of
information~\cite{schumacherQuantumCoding1995}.
A \textit{Quantum gate} is an operator that acts on the state of qubits.
We formally introduce quantum circuits in Section~\ref{sec:quantum_computing}.
%Together they form a graphical
%description of the quantum mechanical operators that act on states.

Quantum circuits provide a formal description of quantum
algorithms, but their practical realization on existing hardware
remains constrained.
The current generation of quantum computers has limited practical use,
as they are part of the \textit{noisy intermediate scale
quantum era (NISQ)}~\cite{preskill2018nisq}.
In particular, the duration in which a quantum device is stable and
can reliably process information ---called \textit{coherence time}---
is limited and restrict the available physical qubits.
This is due to thermal noise, qubit error rate, gate
fidelity, and measurement error~\cite{gottesman2009qec}.
Despite efforts to improve quantum hardware,
mitigation strategies, such as quantum error correction
and quantum circuit optimization,
are required to allow the near-term
usage of quantum computers~\cite{karuppasamy2025qcsurvey,
campbellRoadsFaulttolerantUniversal2017}.

\textit{Quantum circuit optimization} reduces the demand for
resources of quantum algorithms to address these hardware
limitations~\cite{karuppasamy2025qcsurvey, yan2024qcsurvey}.
We distinguish two classes of quantum circuit optimization:
architecture-independent and architecture-dependent.
\textit{Architecture-independent} targets the reduction of noisy
gates and circuit depth.
The reduction of noisy
gates is significant because they are
challenging to implement on current architectures and introduce
substantial error correction overhead~\cite{fowlerSurfaceCodesPractical2012}.
Reducing the circuit depth is linked to speeding up the execution
time of a quantum circuit.
\textit{Architecture-dependent} optimization takes hardware
characteristics into account and maps a quantum circuit to a
specific quantum device.
In particular for superconducting architectures with a grid-based
topology, architecture-aware optimization considers qubit
connectivity and surface codes~\cite{yan2024qcsurvey}.
Qubit connectivity is the physical constraint on qubit interactions.
Surface codes are the quantum error correction algorithms for different gates.

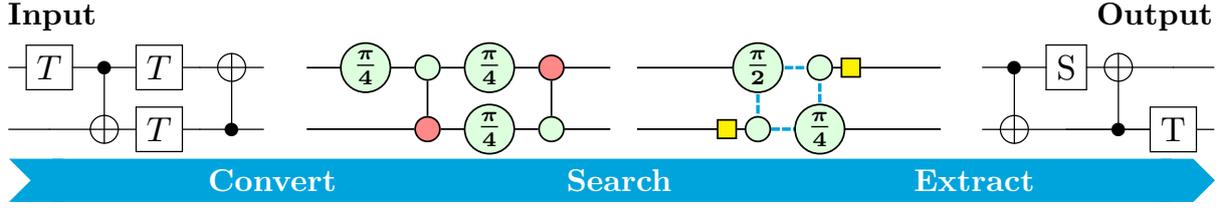
\begin{figure*}[ht!]
  \centering
  \resizebox{\textwidth}{!}{
    \begin{tikzpicture}
	\begin{pgfonlayer}{nodelayer}
        \node[scale=1.1] (A) at (0,0) {
          $\Qcircuit @C=0.5em @R=.5em @!R {
            & \gate{T}  & \ctrl{1}  & \gate{T}  & \qw & \targ    & \qw \\
            & \qw       & \targ     & \gate{T}  & \qw & \ctrl{-1} & \qw}$ 
    };
        \node[align=left] (a) at (-1.9, 2) {
            \textbf{Input}
    };
        \node (aa) at (-3.35, -2) {
    };

    \node[scale=1.5] (B) at (8,0) {
      \begin{tikzpicture}
    \begin{pgfonlayer}{nodelayer}
        \node [style=none] (0) at (0.00, 0.00) {};
        \node [style=none] (1) at (0.00, -1.00) {};
        \node [style=Z phase dot] (2) at (1.00, 0.00) {$\frac{\pi}{4}$};
        \node [style=X dot] (3) at (2.00, -1.00) {};
        \node [style=Z dot] (4) at (2.00, 0.00) {};
        \node [style=Z phase dot] (5) at (3.00, 0.00) {$\frac{\pi}{4}$};
        \node [style=Z phase dot] (6) at (3.00, -1.00) {$\frac{\pi}{4}$};
        \node [style=X dot] (7) at (4.00, 0.00) {};
        \node [style=Z dot] (8) at (4.00, -1.00) {};
        \node [style=none] (9) at (5.00, 0.00) {};
        \node [style=none] (10) at (5.00, -1.00) {};
    \end{pgfonlayer}
    \begin{pgfonlayer}{edgelayer}
        \draw (0) to (2);
        \draw (1) to (3);
        \draw (2) to (4);
        \draw (3) to (4);
        \draw (3) to (6);
        \draw (4) to (5);
        \draw (5) to (7);
        \draw (6) to (8);
        \draw (7) to (8);
        \draw (7) to (9);
        \draw (8) to (10);
    \end{pgfonlayer}
\end{tikzpicture}
    };

        \node[scale=1.5] (C) at (16,0) {
      \begin{tikzpicture}
    \begin{pgfonlayer}{nodelayer}
        \node [style=none] (0) at (0.0, 0.00) {};
        \node [style=none] (1) at (0.0, -1.00) {};
        \node [style=Z phase dot] (2) at (2.00, 0.00) {$\frac{\pi}{2}$};
        \node [style=Z dot] (3) at (2.00, -1.00) {};
        \node [style=Z phase dot] (6) at (3.00, -1.00) {$\frac{\pi}{4}$};
        \node [style=Z dot] (7) at (3.00, 0.00) {};
        \node [style=none] (9) at (5, 0.00) {};
        \node [style=none] (10) at (5, -1.00) {};
        \node [style=hadamard] (11) at (1.50, -1.00) {};
        \node [style=hadamard] (12) at (3.50, 0.00) {};
    \end{pgfonlayer}
    \begin{pgfonlayer}{edgelayer}
        \draw (0) to (2);
        \draw (1) to (3);
        \draw [style=hadamard edge] (2) to (3);
        \draw [style=hadamard edge] (2) to (7);
        \draw [style=hadamard edge] (3) to (6);
        \draw [style=hadamard edge] (6) to (7);
        \draw (6) to (10);
        \draw (7) to (9);
    \end{pgfonlayer}
\end{tikzpicture}
    };
        \node[scale=1.1] (D) at (23.3,0){
          $\Qcircuit @C=0.5em @R=.5em @!R {
	 	 & \ctrl{1}& \gate{\mathrm{S}} & \targ&  \qw   & \qw\\
	 	 & \targ   & \qw               & \ctrl{-1}\qw  & \gate{\mathrm{T}}\qw  &\qw \\}$
    };

        \node (d)[align=left] at (24.8, 2) {
            \textbf{Output}
    };
        \node (dd) at (26.4, -2) {
    };

    \draw[{Triangle Cap[reversed]}-{Triangle Cap},line width=15pt, color=luxembourg blue] (aa) -- (dd);
    \node[color=white] (c) at (3.3, -2) {\textbf{Convert}};
    \node[color=white] (s) at (11.7, -2) {\textbf{Search}};
    \node[color=white] (e) at (20.3, -2) {\textbf{Extract}};
	\end{pgfonlayer}
\end{tikzpicture}
  }
  \caption{Pipeline for ZX-based quantum circuit optimization.}
  \label{fig:optimization}
\end{figure*}

Conventional quantum circuit-based optimization
techniques rely on gate cancellation and gate permutation
rules~\cite{itoko2020gatecommutation}.
%However, several drawbacks emerge from the gate set dependence of the
%quantum circuit representation
However, it is necessary to prove that each simplification rule is
semantics-preserving with respect to the chosen gate set.
Furthermore, multiple rules are often required to capture the same
underlying principle (e.g., two rules per single-qubit gate to commute
through a CNOT gate).

%For example, the commutation of a single-qubit
%gate and CNOT gate adds two unique rules per single-qubit gate of the
%choosen gate set.

ZX-calculus emerged as an
alternative quantum circuit simplification framework beyond the traditional gate
model~\cite{coeckeInteractingQuantumObservables2008} (
Section~\ref{sec:zx_calculus}).
ZX-calculus
expresses the computation as a gate set independent graph called ZX-diagram.
ZX-diagrams are
always composed of the same few elementary building blocks.
%Every quantum circuit can be translated into a ZX
%diagram, but the reverse translation is known as the \textit{circuit
% extraction problem}~\cite{backensThereBackAgain2020}~(
%Section~\ref{sec:circuit_extract}).

ZX-based optimization techniques take advantage of the small and
semantics-preserving set of rewriting rules offered~(Section~\ref{sec:zxrules}).
ZX-calculus enables reductions beyond the gate level for current and
future quantum devices.
Therefore, a review of ZX-based quantum circuit optimization
techniques is required
to evaluate their practical impact and identify opportunities for
future research.

In sum, the contributions of this survey are as follows:
\begin{enumerate}
  \item This survey addresses multiple research communities that are
    unfamiliar with quantum computing or ZX-calculus by
    establishing their respective core
    principles~(Section~\ref{sec:background}).
  \item Comprehensive literature overview of ZX-based quantum circuit
    optimization that is organized by optimization
    strategies, target metrics, and architectural
    dependence~(Sections~\ref{sec:target_metrics}~and~\ref{sec:taxonomy}).
  \item Identification of key challenges and future research
    directions that can benefit from the diverse background of the
    target communities~(Section~\ref{sec:challenges}).
\end{enumerate}

\section{Background}
\label{sec:background}

\paragraph{Quantum computing}
Quantum computing leverages quantum mechanical principles, such as superposition and entanglement, to process information through the sequential application of unitary operators to quantum states (Section~\ref{sec:fundamentals}).
Quantum gates are unitary operators that act on qubits (Section~\ref{chap1:sec:quantum-gates}).
A quantum circuit is a linear map of qubits that is composed of sequential quantum gates (Section~\ref{chap1:sec:quantum-circuits}).
Quantum circuit optimization improves the resource requirements of quantum algorithms for current NISQ-era and future fault-tolerant quantum devices (Section~\ref{chap1:sec:quantum-optimization}).

\paragraph{ZX-calculus}
A ZX-diagram is tensor network composed by generators that can implement the linear map of qubits for every quantum circuit (Section~\ref{chap2:sec:zx-fundamentals}).
Phase gadgets, phase-polynomials, Pauli gadgets and spider nests are macroscopic structures commonly found in ZX-diagrams that permit efficient optimization (Section~\ref{chap1:sec:macroscopic-structures}).
The rewriting rules of ZX-calculus allow for semantics-preserving transformations of ZX-diagrams (Section~\ref{sec:zxrules}).
Circuit extraction from ZX-diagrams is computationally expensive, and the specified extraction algorithm determines the characteristics of the resulting quantum circuit (Section~\ref{chap1:sec:circuit-extraction-problem}).
The presence of causal flow, general flow, or Pauli flow~\cite{duncanGraphtheoreticSimplificationQuantum2020,backensThereBackAgain2020,simmons2021Measurement}is sufficient to assert a deterministic behavior necessary for polynomial-time circuit extraction (Section~\ref{chap1:sec:flow}).

\section{Quantum Computing}
\label{sec:quantum_computing}
This section introduces a concise and minimal summary of quantum computing necessary to understand the connection between quantum circuits and ZX-calculus. 
This short introduction follows the de facto reference textbook for quantum computing and information theory by Nielsen and Chuang~\cite{nielsenQuantumComputationQuantum2010a}.
Quantum mechanical descriptions and examples are taken from Griffiths~\cite{griffithsIntroductionQuantumMechanics2005}.

\subsection{Fundamentals}
\label{sec:fundamentals}
\subsubsection{Key Concepts}
Quantum systems are described by a complex \emph{Hilbert space}, an inner product space that is complete in the induced norm.
Physical pure quantum states are represented in a complex Hilbert space by rays, an equivalence class of vectors that differ by the multiplication of a nonzero complex scalar.
Vectors that only vary by a global phase characterize the same quantum state.

The \emph{Dirac notation} is a widely used description for the linear algebra of complex vector spaces, such as Hilbert spaces, encountered in quantum mechanics~\cite{diracNewNotationQuantum1939}. 
In the Dirac notation, a quantum state is named ket $\ket{\phi}$ and defined up to a complex phase as a normalized Hilbert space vector.
A bra $\bra{\psi}$  is defined as the Hermitian conjugate of a ket $\ket{\psi}$, such that $\bra{\psi} = \left(\ket{\psi}\right)^{\dagger}$.
For Hilbert spaces of finite dimensions, this corresponds to the conjugate transpose of a ket vector.
The inner product $\braket{\phi}{\psi}$ gives the probability amplitude of a quantum system that originates in $\ket{\psi}$ can be found in state $\ket{\phi}$ upon measurement. 
The Born rule states that the probability to obtain outcome $\ket{\phi}$ when measuring state $\ket{\psi}$ is given by $|\braket{\phi}{\psi}|^2$~\cite{born1926quantenmechanik}.
All states throughout this section are assumed to be normalized vectors ($\braket{\phi}{\phi} = 1$) to ensure that the total probability from measurement outcomes equals one.

A key concept in quantum mechanics is \emph{superposition}.
It signifies that a quantum state can be expressed as a linear combination of basis states with coefficients being complex valued probability amplitudes.
If $\ket{\phi}$ and $\ket{\psi}$ form an orthonormal basis, then $\abs{\alpha}^{2}$ and $\abs{\beta}^{2}$ are the probabilities to obtain the states $\ket{\phi}$ and $\ket{\psi}$ when measuring $\ket{\Psi}$.
\begin{align*}
  \ket{\Psi} & = \alpha \ket{\phi} + \beta \ket{\psi} \\
  1 &= \abs{\alpha}^{2} + \abs{\beta}^{2}
\end{align*}\\

The tensor product composes the joint state between quantum systems.
For two quantum systems with states $\ket{\phi}$ and $\ket{\psi}$, their joint state $\ket{\Psi}$ is given by the tensor product, such that $\ket{\Psi} = \ket{\phi}\otimes\ket{\psi}$.

\emph{Entanglement} is another key concept in quantum mechanics.
A composite state is entangled if it cannot be decomposed as a tensor product between states of the individual subsystems.

\emph{Operators} $\hat{U}$ are linear maps that act on quantum states.
The application of an operator $\hat{U}$ onto a state $\ket{\phi}$ result in a new state $\ket{\phi_{U}}$, such that $\ket{\phi_{U}}=\hat{U}\ket{\phi}$.
Generally, operators do not commute $UV\neq VU$; hence the order in which the operators act on a state remains relevant.

Quantum operators are linear maps that act onto the Hilbert space.
\sloppy For two states $\ket{\phi}$, $\ket{\psi}$ with complex coefficients $a,b$, the operator satisfies $\hat{U}\left(a\ket{\phi} + b\ket{\psi}\right) = a\hat{U}\ket{\phi} +b\hat{U}\ket{\psi}$ holds.
Furthermore, the time evolution of a quantum state in a closed system is described by \emph{unitary operators}, such that $\hat{U}^{\dagger}\hat{U}=\hat{1}$. 
This restriction is required because only unitary operations preserve the norm of a quantum state vector and consequently the total probability under time evolution. 

\emph{Hermitian operators} remain unchanged under hermitian conjugation, such that $\hat{U}^{\dagger} = \hat{U}$.
These operators represent physical observables and admit real eigenvalues.
Unitary operators that are also Hermitian satisfy $\hat{U}^{2}=\hat{1}$ and recover the original state after being applied twice.

\begin{table*}[t!]
  \centering
  \begin{tabularx}{\textwidth}{lll}
    \toprule
    \emph{Pauli Basis} & \emph{Matrix} &
    \emph{Eigenstates} \\
    \midrule
    \( Z \)-Basis (Computational) &
    \( Z =
      \begin{bmatrix}1 & 0 \\ 0 & -1
    \end{bmatrix} \) &
    \( \ket{0} =
      \begin{bmatrix}1 \\ 0
      \end{bmatrix}, \quad \ket{1} =
      \begin{bmatrix}0 \\ 1
    \end{bmatrix} \)  \\\\
    \( X \)-Basis (Hadamard) &
    \( X =
      \begin{bmatrix}0 & 1 \\ 1 & 0
    \end{bmatrix} \) &
    \( \ket{+} = \frac{\ket{0} + \ket{1}}{\sqrt{2}}, \quad \ket{-} =
    \frac{\ket{0} - \ket{1}}{\sqrt{2}} \) \\\\
    \( Y \)-Basis &
    \( Y =
      \begin{bmatrix}0 & -i \\ i & 0
    \end{bmatrix} \) &
    \( \ket{+i} = \frac{\ket{0} + i\ket{1}}{\sqrt{2}}, \quad \ket{-i}
    = \frac{\ket{0} - i\ket{1}}{\sqrt{2}} \)  \\
    \bottomrule
  \end{tabularx}
  \caption{Pauli Matrices and their Eigenstates as a superposition of
  the computational basis.}
  \label{tab:paulibases}
\end{table*}

\subsubsection{Qubits}
The basic unit of information in quantum computing is the \emph{qubit}~\cite{schumacherQuantumCoding1995}.
In contrast to a classical bit that can only represent states 0 and 1, qubits can be in a superposition of the computational basis states $\ket{0}$ and $\ket{1}$.
A basis is a set of orthonormal states that can expressed any state of the Hilbert space as the linear combination of these state vectors.

\begin{figure}[h!]
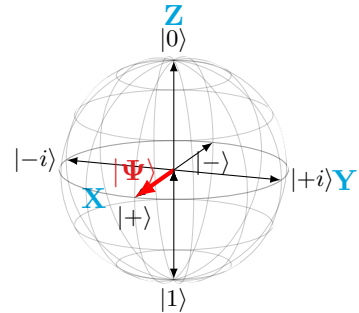

  \centering
  \begin{blochsphere}[radius=1.5 cm,tilt=15,rotation=-20,
    color=white, opacity = 0.1, ball=circle, ]
    \drawBallGrid[style={opacity=0.2}]{30}{30}

    \labelLatLon{up}{90}{0};
    \labelLatLon{down}{-90}{90};
    \labelLatLon{left}{-180}{0};
    \labelLatLon{right}{0}{0};
    \labelLatLon{front}{0}{90};
    \labelLatLon{behind}{0}{-90};
    \labelLatLon{state}{0}{90};

    \draw[-latex] (0,0) -- (up) node[above,inner sep=.5mm] at (up)
    {\footnotesize $\ket{0}$};
    \draw[-latex] (0,0) -- (down) node[below,inner sep=.5mm] at
    (down) {\footnotesize $\ket{1}$};

    \draw[-latex] (0,0) -- (left) node[left,inner sep=.5mm] at (left)
    {\footnotesize $\ket{-i}$};
    \draw[-latex] (0,0) -- (right) node[right,inner sep=.5mm] at
    (right) {\footnotesize $\ket{+i}$};

    \draw[-latex] (0,0) -- (front) node[below,inner sep=.5mm] at
    (front) {\footnotesize $\ket{+}$};
    \draw[red,ultra thick, -latex] (0,0) -- (front) node[luxembourg red,
    above,inner sep=.7mm] at (front) {$\mathbf{\ket{\Psi}}$};
    \draw[-latex] (0,0) -- (behind) node[below,inner sep=.5mm] at
    (behind) {\footnotesize $\ket{-}$};

    \draw[-latex] (0,0) node[above,inner sep=4.0mm] at (up)
    {\color{luxembourg blue}\textbf{Z}};
    \draw[-latex] (0,0) node[right,inner sep=6.0mm] at (right)
    {\color{luxembourg blue}\textbf{Y}};
    \draw[-latex] (0,0) node[left,inner sep=3.0mm] at (front)
    {\color{luxembourg blue}\textbf{X}};

  \end{blochsphere}

  \caption{Bloch sphere.}
  \label{fig:blochsphere}
\end{figure}

Besides the computational basis, \emph{Pauli bases} are frequently encountered in quantum computing (Table~\ref{tab:paulibases}).
An intuitive tool to visualize pure single-qubit states up to a global phase is the Bloch sphere~\cite{Bloch1946}.
Figure~\ref{fig:blochsphere} shows a \emph{Bloch sphere} with the state $\ket{\Psi} = \frac{\ket{0} + \ket{1}}{\sqrt{2}} = \ket{+}$ indicated in red.
The different axes correspond to the Pauli bases -X ($\ket{-}$, $\ket{+}$), -Y ($\ket{-i}$, $\ket{+i}$) and -Z ($\ket{0}$, $\ket{1}$) bases.
The probability of a measurement outcome depends on the basis it is measured in as stated by the Born rule.
A measurement of $\ket{\Psi}$ on the basis of Z would result in $\ket{0}$ and $\ket{1}$ with a probability of 50\% for each outcome. 
If the same state $\ket{\Psi}$ is measured in the X-basis instead, the final state would always be $\ket{+}$.

\emph{Entanglement} connects the state of multiple qubits such that the formed composite system can not be separated by as the tensor products of its individual systems~\cite{einsteinCanQuantumMechanicalDescription1935}. 
For example, the composite state of two qubits in the computational basis is described by the tensor product of the computational basis states, thus $\mathbf{\ket{00} = \ket{0} \otimes \ket{0}}$.

Other prominent examples of entanglement are the \emph{Bell states}~\cite{bell1964}.
Bell states are maximally entangled two-qubit states.
Maximally entangled qubits are perfectly correlated, and the measurement of one qubit will result in correlated outcomes regardless their distance.
Highly entangled qubits are an essential resource for quantum error correction~\cite{knill1997,fowlerSurfaceCodesPractical2012}.

\emph{Coherence time} is the timespan in which a quantum system maintains its phase coherence.
Over time, accumulated environmental noise, e.g. from heat, forces decoherence of the quantum states, limiting the duration in which quantum systems can reliably process information.

\subsection{Quantum Circuits}
The quantum circuit model is one way to express quantum computation. 
Quantum circuits are graphical representations of the manipulation of quantum-mechanical states (qubits) by operators (quantum gates)~\cite{barencoElementaryGatesQuantum1995a}.
Like classical electrical circuits, quantum circuits implement the semantics and control flow of a program. 
Individual operations are performed by quantum gates.
%The literature differentiates between random and structured
%quantum circuits.
%Randomly generated circuits are typically generated with different
%weights for the various quantum gates and do not represent
%a real-world application.
%Structured circuits are quantum programs that solve real-world
%problems.

\subsubsection{Quantum Gates}
\label{chap1:sec:quantum-gates}
\begin{figure*}[ht!]
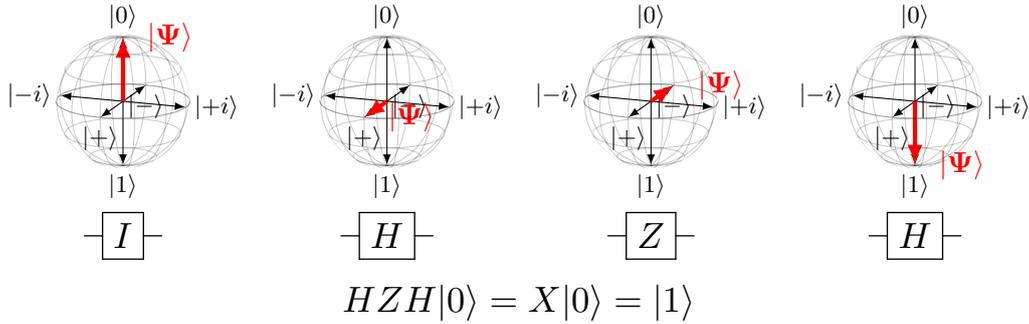

  \centering
  \renewcommand{\tabcolsep}{1pt}
  \resizebox{\textwidth}{!}{
    \begin{tabular}{cccc}
      \blochstate{up} & \blochstate{front} & \blochstate{behind} &
      \blochstate{down} \\
      \scalebox{1.25}{$\qcid$} & \scalebox{1.25}{$\qch$} &
      \scalebox{1.25}{$\qcz$} &  \scalebox{1.25}{$\qch$}\\[1em]
      \multicolumn{4}{c}{
        \scalebox{1.25}{$ HZH \ket{0} = X \ket{0} = \ket{1}$}
      }
  \end{tabular}}
  \caption{Application sequence (left to right) of single-qubit gates
  starting at $\ket{0}$.}
  \label{fig:singlequbitrotations}
\end{figure*}

The elementary building blocks of quantum circuits are quantum gates.
Like their classical counterparts, the logic gate in digital computers, quantum gates implement elementary operations onto a state. 
In contrast to classical logic gates, unitary quantum gates are \emph{reversible}. 
The reversibility of computation mandates that the original states are always recoverable from the final states~\cite{shendeReversibleLogicCircuit2003}.
Certain quantum gates have classical analogs; the quantum $X$ gate is the quantum equivalent of the classical NOT gate. 
Other gates are unique for the quantum computing paradigm without a classical counterpart; unique quantum gates include the Hadamard $H$ and $Z$ gate.

From a quantum-mechanical perspective, all quantum gates are unitary operators that can be represented as unitary matrices. 
Quantum gates can operate on one or multiple qubits. 
Single-qubit gates can be visualized as rotations on the Bloch sphere.
Figure~\ref{fig:singlequbitrotations} highlights the sequential application of the $HZH$ operators to the $\ket{0}$ state, acting effectively as the $X$ gate. 
The $Z$, $X$ and $H$ gates reappear later in the form of the elementary building blocks of ZX-calculus. 
If a quantum gate is both a unitary and a Hermitian operator, applying the same gate twice recovers the original state.

Quantum computers can implement universal gate sets to perform every computation. 
Although not universal, the Clifford gate set is a popular choice as a consequence of the Gottesman-Knill theorem, which states that Clifford computations can be efficiently simulated classically~\cite{gottesman1998heisenbergrepresentationquantumcomputers,aaronsonImprovedSimulationStabilizer2004}.
Although there are an infinite number of universal gate sets, in practice the \emph{Clifford+T} set is chosen due to its importance to fault-tolerant compilation where the cost of non-Clifford gates, such as the T-gate, are higher than Clifford gates~\cite{fowlerSurfaceCodesPractical2012}. 
Table~\ref{tab:quantum_gates} lists the gates that compose the Clifford + T-gate set with the corresponding unitary matrices and the quantum circuit notation.

As we saw before, quantum gates are unitary operators that do not necessarily commute. 
Therefore, different universal gate sets admit different gate commutation rules (e.g., Clifford+T or Toffoli).
This is of particular importance, for the field of rule-based quantum circuit optimization~\cite{gottesman1998heisenbergrepresentationquantumcomputers,itoko2020gatecommutation}.

\begin{table*}[pht!]
  \centering
  \resizebox{\textwidth}{!}{
    $
    \begin{array}{lccccc}
      \toprule
      \emph{Gate} & \emph{Unitary Matrix} &
      \emph{quantum circuit} & \emph{ZX-Calculus} &  \emph{Hermitian} &
      \emph{Single-Qubit} \\\midrule
      \text{Identity} &  \identity =
      \begin{bmatrix} 1 & 0 \\ 0 & 1
      \end{bmatrix}  &  \scalebox{1.5}{$\qcid$} &
      \scalebox{1.5}{
        \begin{ZX}[ampersand replacement=\&] \zxNone{} \ar[r] \& \zxNone{} \rar \&[\zxWCol] \zxNone{} \ar[r] \& \zxNone{}
      \end{ZX} } & \yes & \yes\\\\
      \text{Pauli-X} &  X =
      \begin{bmatrix} 0 & 1 \\ 1 & 0
      \end{bmatrix}  & \scalebox{1.5}{$\qcx$} &
      \scalebox{1.5}{
        \begin{ZX}[ampersand replacement=\&] \zxN{} \rar \&[\zxwCol] \zxX{\pi} \rar \&[\zxwCol] \zxN{}
      \end{ZX}} & \yes & \yes \\\\
      \text{Pauli-Y}  &  Y=
      \begin{bmatrix} 0 & -i \\ i & 0
      \end{bmatrix}  & \scalebox{1.5}{$\qcy$} &
      \scalebox{1.5}{
        \begin{ZX}[ampersand replacement=\&] \zxN{} \rar \&[\zxwCol] \zxZ{\pi} \rar \&[\zxwCol]
          \zxX{\pi} \rar \&[\zxwCol] \zxN{}
      \end{ZX}} & \yes & \yes \\\\
      \text{Pauli-Z} &  Z =
      \begin{bmatrix} 1 & 0 \\ 0 & -1
      \end{bmatrix}  & \scalebox{1.5}{$\qcz$} &
      \scalebox{1.5}{
        \begin{ZX}[ampersand replacement=\&] \zxN{} \rar \&[\zxwCol] \zxZ{\pi} \rar \&[\zxwCol] \zxN{}
      \end{ZX}} &  \yes & \yes \\\\
      \text{Phase}  &  S=
      \begin{bmatrix} 1 & 0 \\ 0 & i
      \end{bmatrix}  & \scalebox{1.5}{$\qcs$} &
      \scalebox{1.5}{
        \begin{ZX}[ampersand replacement=\&] \zxN{} \rar \&[\zxwCol] \zxZ{\frac{\pi}{2}} \rar
          \&[\zxwCol] \zxN{}
      \end{ZX}} & \no & \yes \\\\
      \text{S}^\dagger  &  S^\dagger  =
      \begin{bmatrix} 1 & 0 \\ 0 & -i
      \end{bmatrix}   & \scalebox{1.5}{$\qcsd$} &
      \scalebox{1.5}{
        \begin{ZX}[ampersand replacement=\&] \zxN{} \rar \&[\zxwCol] \zxZ{-\frac{\pi}{2}} \rar
          \&[\zxwCol] \zxN{}
      \end{ZX}} & \no & \yes \\\\
      \text{T} &  T =
      \begin{bmatrix} 1 & 0 \\ 0 & e^{i\pi/4}
      \end{bmatrix}  & \scalebox{1.5}{$\qct$} &
      \scalebox{1.5}{
        \begin{ZX}[ampersand replacement=\&] \zxN{} \rar \&[\zxwCol] \zxZ{\frac{\pi}{4}} \rar
          \&[\zxwCol] \zxN{}
      \end{ZX}} & \no & \yes \\\\
      \text{T}^\dagger &  T^\dagger =
      \begin{bmatrix} 1 & 0 \\ 0 & e^{-i\pi/4}
      \end{bmatrix}   & \scalebox{1.5}{$\qctd$} &
      \scalebox{1.5}{
        \begin{ZX}[ampersand replacement=\&] \zxN{} \rar \&[\zxwCol] \zxZ{-\frac{\pi}{4}} \rar
          \&[\zxwCol] \zxN{}
      \end{ZX}} & \no & \yes \\\\
      \text{Hadamard}  &  H = \frac{1}{\sqrt{2}}
      \begin{bmatrix} 1 & 1 \\ 1 & -1
      \end{bmatrix}  & \scalebox{1.5}{$\qch$} &
      \scalebox{1.5}{
        \begin{ZX}[ampersand replacement=\&] \zxN{} \rar \&[\zxwCol] \zxH{} \rar \&[\zxwCol] \zxN{}
      \end{ZX}} & \yes & \yes \\\\
      \text{CNOT}  &  CX =
      \begin{bmatrix} 1 & 0 & 0 & 0 \\ 0 & 1 & 0 & 0 \\ 0 & 1 & 0 & 1
        \\ 0 & 0 & 1 & 0
      \end{bmatrix}  & \vcenter{\hbox{\scalebox{1.5}{$\qccx$}}} &
      \scalebox{1.5}{$
        \begin{ZX}[sep=0.75em,inner sep=1.5em, row sep=0.75em, baseline={([yshift=0.07em]current bounding box.center)}, ampersand replacement=\&]\zxN{} \rar \& [\zxwCol] \zxZ{}  \dar
          \rar \& [\zxwCol]
          \zxN{} \\ \zxN{} \rar \& \zxX{} \rar \& [\zxwCol] \zxN{} \\
      \end{ZX}$} & \vcenter{\hbox{\shortstack{\yes
      \\ \text{\phantom{(2-qubit)}}}}}  & \vcenter{\hbox{\shortstack{\no \\\text{(2-qubit)}}}} \\\\\midrule\\
      \text{Controlled Z}  & CZ =
      \vcenter{\hbox{%
          $
          \begin{bmatrix}
            1 & 0 & 0 & 0 \\
            0 & 1 & 0 & 0 \\
            0 & 0 & 1 & 0\\
            0 & 0 & 0 & -1\\
      \end{bmatrix}$}} &
      \vcenter{\hbox{\scalebox{1.5}{$\qccz$}}} &
      \scalebox{1.5}{
        $
        \begin{ZX}[sep=0.75em, row sep=1em, baseline={([yshift=0.07em]current bounding box.center)}, ampersand replacement=\&]\zxN{} \rar \& [\zxwCol] \zxZ{}  \dar[H]
          \rar \& [\zxwCol]
          \zxN{} \\ \zxN{} \rar \& \zxZ{} \rar \& [\zxwCol] \zxN{} \\
      \end{ZX}$} &\vcenter{\hbox{\shortstack{\yes\\ \phantom{\text{(2-qubit)}}}}} & \vcenter{\hbox{\shortstack{\no\\ \text{ (2-qubit)}}}} \\\\\bottomrule
    \end{array}$
  }
  \caption{Quantum gates that compose the universal Clifford+T gate set and the controlled Z (CZ) gate.}
  \label{tab:quantum_gates}
\end{table*}

\subsubsection{Quantum Circuits}
\label{chap1:sec:quantum-circuits}
To process information, quantum circuits combine an application sequence of quantum gates on qubits. 
Quantum circuits are a linear map of qubits.

They are composed of matrix and tensor products of operators.
The matrix product is used for the sequential composition of gates along the program flow.
The tensor product describes the composite system that arise from parallel gate application.
As not all quantum gates commute, the exact sequence needs to be preserved upon composition of the full operator such that the semantics are preserved.

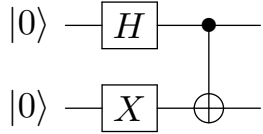
\begin{figure}[h!]
  \centering
  \scalebox{1.25}{$\Qcircuit @C=1em @R=1em {
      \ket{0} & & \gate{H}   & \ctrl{1} & \qw \\
      \ket{0} & & \gate{X}        & \targ    & \qw
  }$}
  \caption{Generating circuit of the $\ket{\Psi^{+}}$ Bell state.}
  \label{fig:examplecircuit}
\end{figure}

Figure~\ref{fig:examplecircuit} shows a quantum circuit that generates the $\ket{\Psi^{+}}$ Bell state.
The control flow is from left to right.
Starting from $\ket{00}$, an $H$ gate is applied on the first qubit and a $X$ gate on the second qubit, resulting in $\left(\frac{\ket{01}+\ket{11}}{\sqrt{2}}\right)$.
Lastly, a CNOT gate is placed giving rise to the $\ket{\Psi^{+}}$ Bell state:
\begin{equation*}
  CX\dot(H\otimes X)\ket{00} = \frac{1}{\sqrt{2}}(\ket{01}+\ket{10})
  = \ket{\Psi^{+}}
\end{equation*}

The familiar look of quantum circuits of classical computing should not deceive one from the fact that quantum computing is a fundamentally different paradigm that enforces different constraints not known by classical computation.
The \emph{no-cloning theorem} states that it is impossible to create a perfect and independent copy of an arbitrary unknown quantum state~\cite{Wootters1982,Dieks1982}.
As a direct consequence, quantum computing cannot use classical error correction codes that are based on the repetition of direct copying.
Different quantum error correction schemes exist that encode information into entangled states (e.g., Shor~\cite{shor1995qec}, Steane~\cite{steane1996qec}, and surface codes~\cite{bravyi1998qec}).

\subsubsection{Quantum Circuit Optimization}
\label{chap1:sec:quantum-optimization}
Current generation quantum hardware possesses several resource restrictions, namely physical qubit availability and limited coherence time.
Quantum circuit optimization addresses these limitations and can be broadly divided into two categories~\cite{karuppasamy2025qcsurvey}.

\emph{Architecture-independent optimization} focuses on hardware-agnostic strategies at the logical qubit level without any consideration of their physical implementation.
Logical qubits are error corrected qubits composed of several physical qubits.
Physical qubits are the actual hardware elements. 
They are prone to noise and are not error corrected.
With the aid of quantum error correction, multiple physical qubits form a logical qubit.

Typical optimization targets for NISQ-era quantum devices multi-qubit gates (e.g., CNOT, CZ and SWAP) due to their high implementation cost and connectivity constraints~\cite{preskill2018nisq}, while the cost of fault-tolerant architectures is driven by T-gates~\cite{gottesman2009qec}.

%One example to implement error correction are surface codes, additional quantum circuit logic, that add more overhead for different type of quantum gates~\cite{fowlerSurfaceCodesPractical2012}.
Decreasing the circuit depth reduces the execution time of a quantum circuit. 
Speeding up the execution time is especially important if the unoptimized circuit cannot be executed during the coherence time.
Furthermore, quantum circuit synthesis aims to decompose arbitrary unitary operations into an optimal sequence of quantum gates~\cite{yan2024qcsurvey}.

\emph{Architecture-dependent optimization} concentrates on hardware specific characteristics, primarily a circuit's qubit connectivity (which qubits can directly interact with each other), gate fidelity (accuracy of a gate's operation), circuit fidelity (accuracy of a full circuit's unitary operator), and error rate (probability of erroneous state changes)~\cite{zhu2025quantumcompilerdesignqubit}. 
Multi-qubit gates can only operate on connected qubits. 
Extra operations, such as SWAP gates, facilitate connections between qubits.
Routing optimizes maps a quantum circuit onto a quantum device and minimizes the overhead introduced by the inserted SWAP gates.

A quantum architecture encodes logical qubits using many physical qubits according to an quantum error correction scheme. 
Additionally, architecture-aware synthesis decomposes and aligns parts of the quantum circuit with respect to the connectivity and limitations of a given quantum architecture~\cite{yan2024qcsurvey}.

\section{ZX-Calculus}
\label{sec:zx_calculus}
ZX-calculus is a diagrammatic reasoning framework for quantum circuits.
Both represent linear maps of qubits, but ZX-calculus offers a compact representation of complex quantum programs and admit a sound and complete set of rewriting rules.
Originally introduced in 2008 by Coecke and Ross~\cite{coeckeInteractingQuantumObservables2008}, ZX-calculus is gaining popularity in the field of quantum circuit optimization with further applications in quantum circuit verification.
This section follows the book-sized introduction of Coecke and Ross~\cite{coeckePicturingQuantumProcesses2017} and the shorter introductory paper van de Wetering~\cite{weteringZXcalculusWorkingQuantum2020}.

%dixon2013opengraph
\subsection{Fundamentals}
\label{chap2:sec:zx-fundamentals}
The objects in ZX-calculus are referred to as ZX-diagrams, a type tensor networks that represent the linear map among qubits.
The generators of ZX-diagrams are \emph{spiders}.
A spider represents a tensor that operates on qubits in the Z-basis $\left\{\ket{0}, \ket{1} \right\}$ (green) or X-basis $\left\{\ket{-}, \ket{+} \right\}$ (red).
Spiders have $n$ inputs, $m$ outputs, and carry a phase of $\alpha$.
The following figure visualizes a Z- and X-spider with $n$ inputs and $m$ outputs and their respective unitary operator in Dirac notation.
\begin{align*}
  \scalebox{1.25}{
    \begin{ZX}  \leftManyDots{n} \zxZ{\alpha} \rightManyDots{m}
  \end{ZX}} &=\ket{0, ...,0}\bra{0,...,0}  \\
  &+ e^{i\alpha}\ket{1, ...,1}\bra{1,...,1} \\
  \scalebox{1.25}{
    \begin{ZX}  \leftManyDots{n} \zxX{\alpha} \rightManyDots{m}
  \end{ZX}} &= \ket{+, ...,+}\bra{+,...,+} \\
  &+ e^{i\alpha}\ket{-,...,-}\bra{-,...,-}\\
\end{align*}

\emph{Wires} connect the output of one spider with the input of another spider. 
They implement the identity linear map on a qubit and leave the quantum state unchanged:
\begin{equation*}
  \scalebox{1.5}{
    \begin{ZX}[ampersand replacement=\&] \zxN{} \rar \&[\zxWCol] \zxN{}
  \end{ZX}}~=~\ket{0}\bra{0}  + \ket{1}\bra{1}
\end{equation*}
Phase-less spiders with single input and output wire, implement the identity linear map and therefore act like wires.

The Hadamard gate can be represented by either a yellow box connected by wires or a blue dotted line between spiders.
The blue dotted line appears in the graph-formalism and is called a \emph{Hadamard edge / wire}.
Euler decomposition, known as the \emph{Hadamard rule (hd)}, splits up a Hadamard wire into a sequence of Z and X spiders~\cite{duncanGraphStatesNecessity2009}.
\begin{equation*}
  \scalebox{1.5}{
    \begin{ZX}[ampersand replacement=\&]
      \zxN{} \rar \&[\zxwCol] \zxH{} \rar \&[\zxwCol] \zxN{}
  \end{ZX}}~=~
  \scalebox{1.5}{
      \begin{ZX}[ampersand replacement=\&] \zxN{} \ar[r] \& \zxZ{} \ar[r, draw=luxembourg blue, thick, densely dotted] \&[\zxWCol]  \zxZ{} \ar[r] \& \zxN{} \end{ZX}}~=
  ~\scalebox{1.25}{
    \begin{ZX}[ampersand replacement=\&]
      \zxN{} \ar[r] \& \zxFracZ{\pi}{2} \ar[r] \& \zxFracX{\pi}{2}
      \ar[r] \& \zxFracZ{\pi}{2} \ar[r] \& \zxN{}
  \end{ZX}}
\end{equation*}

A typical ZX-diagram consists of many connected spiders.
Matrix multiplication composes the linear map of sequentially connected spiders and the tensor product composes the linear map between parallel spiders.
ZX-calculus is \emph{universal}, therefore every quantum circuit can be expressed as a ZX-diagram, but the reverse is not trivial and is known as the circuit extraction problem (Section~\ref{chap1:sec:circuit-extraction-problem}).
Two ZX-diagrams are considered equal, if they implement the same linear map up to a global complex phase.

ZX-calculus is especially suited for quantum circuit optimization because it offers a rewriting system that is:
\begin{itemize}
  \item \emph{Sound:} all rules are semantic-preserving and provable correct.
  \item \emph{Compact:} small number of elementary rewriting rules for different fragments, e.g. the scalar-free ZX-calculus with the Clifford fragment shown in Figure~\ref{fig:rules}.
  \item \emph{Complete:} all rewriting rules are derivable from first principles within the ZX-calculus up to arbitrary real phases~\cite{backensZXcalculusCompleteStabilizer2014,backensMakingStabilizerZXcalculus2015,jeandelCompleteAxiomatisationZXCalculus2018,jeandelDiagrammaticReasoningClifford2018}.
\end{itemize}

ZX-calculus demonstrates its effectiveness for quantum circuit optimization in particular, because rather than adhering to the rigid quantum circuit structure with local gate commutation and cancellation rules (e.g., Clifford gates~\cite{gottesman1998heisenbergrepresentationquantumcomputers} and rotation gates~\cite{itoko2020gatecommutation}), it works underlying symmetries and structures of the linear maps~\cite{duncanGraphtheoreticSimplificationQuantum2020}.

%Given the undirected nature of edges and lack of temporal ordering inside ZX-diagrams causes a loss of determinism and give rise to the circuit extraction problem (Section~\ref{chap1:sec:circuit-extraction-problem}).

%The elementary building blocks of quantum circuits are quantum gates and wires. 
%Analogously, the elementary building blocks in the graph-formalism of ZX-calculus---called \emph{generators}---are spiders.
%In contrast to quantum circuits, which can be defined by an infinite amount of universal gate sets, ZX-calculus only contains the same 8 generators irrespective of a quantum circuit's chosen gate set.

%ZX-diagrams are tensor networks generated by spiders.

%The swap generator swaps the spiders on a wire and implements the
%same linear map as the swap gate in the quantum circuit notation.
%\begin{align*}
%  \scalebox{1.25}{
%    \begin{ZX}[ampersand replacement=\&]
%      \zxNoneDouble|{} \ar[r,s,start anchor=north,end anchor=south]
%      \ar[r,s,start
%      anchor=south,end anchor=north] \&[\zxWCol] \zxNoneDouble|{}
%  \end{ZX}} &=~\ket{00}\bra{00}  + \ket{01}\bra{10}\\
%  &+~\ket{10}\bra{01}+ \ket{11}\bra{11}
%\end{align*}
%In ZX-calculus, bent wires depict the Bell state and the Bell effect and
%are known as cup and cap.
%\begin{align*}
%  \scalebox{1.25}{
%    \begin{ZX}
%      \zxNone{} \ar[d,C] \\[\zxWRow]
%      \zxNone{}
%  \end{ZX}}~&=~\ket{00} + \ket{11}\\
%  \scalebox{1.25}{
%    \begin{ZX}
%      \zxNone{} \ar[d,C-] \\[\zxWRow]
%      \zxNone{}
%  \end{ZX}}~&=~\bra{00} + \bra{11} \\
%\end{align*}

Although a spiders phase can be a real number in the general case, restricting the validly assignable phases can be useful to represent specific universal gate sets. 
Spiders with phases that are multiples of $\frac{\pi}{2}$ can implement all Clifford gates. 
Figure~\ref{fig:rules} shows the basic rewriting rules of the scalar-free Clifford ZX-calculus.
A T-gate corresponds to a Z-spider with a phase of $\frac{\pi}{4}$. 
Clifford gates and the T-gate form a universal gate set together.

ZX-diagrams form a dagger compact product and permutation (PROP) where topological transformations, such as bending or moving wires without changing the connectivity, do not change the implemented linear map~\cite{maclane1965prop,coeckeInteractingQuantumObservables2011}.

In this survey, we focus on the graph-formalism of the ZX-calculus..
Every ZX-diagram can be translated into its graph-like representation  (Definition~\ref{def:graph-like}) using the color changing, spider fusion
\begin{definition}{Graph-like ZX-diagrams}
    \label{def:graph-like}
    A ZX-diagram is graph-like if it admits the following characteristics~\cite{duncanGraphtheoreticSimplificationQuantum2020}:
    \begin{enumerate}
        \item All spiders are Z-spiders.
        \item No parallel Hadamard wires or self-loops.
        \item Z-spiders are only connected by Hadamard wires.
        \item All inputs and outputs are connected to a Z-spider. 
        \item Every Z-spider is connected to maximal one input or output.
    \end{enumerate}
\end{definition}

\subsection{Macroscopic Structures}
\label{chap1:sec:macroscopic-structures}
%\begin{itemize}
%    \item \xxx{This section is too abstract}
%    \item \xxx{extend explanations a little bit}
%    \item \xxx{show the difference between a phase gadget and a pauli gadget}
%    \item \xxx{i am not sure that people understand what a parity matrix is}
%\end{itemize}
This section introduces macroscopic structures, larger scale
patterns of generators, commonly
found in ZX-diagrams that are used by some of the surveyed works.
\emph{Phase gadgets} are structures in ZX-diagrams that add a phase to a state~\cite{kissinger2020journal,cowtanPhaseGadgetSynthesis2020}. 
Legs are wires that connect spiders of the same color to a single spider of the opposite color, which subsequently connects to a state.
Figure~\ref{figphasegadget} visualizes a phase gadget with three legs on the left.
Their unitary can be expressed as matrix exponentials $e^{-i\theta\oplus_{i}Z_{i}}$ or $e^{-i\theta\oplus_{i}X_{i}}$.
Figure~\ref{figphasegadget} shows the phase gadget and its decomposition into a CNOT ladder that implements the unitary $e^{-i\theta ZZZ }$.
\emph{Phase polynomials} are a class of circuits that are only composed of CNOT, and phase gates acting in the Z-basis.
These circuits are interesting because they can be fully described by an unitary operator that takes the form of a matrix exponential $e^{-i\frac{\theta}{2}(1-2Px)}$ with $P$ being the circuit's parity table.
Phase polynomials are a sequence of phase gadgets. 
While phase polynomials are based on the parity operator $\oplus$, \emph{spider nests} are based on the and $\wedge$ operator~\cite{munson2021And}. 
\emph{Pauli gadgets} extend the notion of phase gadgets by allowing each leg to connect to spiders that can be of type X, Y or Z.

\begin{figure}[h!]
  \begin{center}
    %\resizebox{\textwidth}{!}{
    \begin{tikzpicture}
	\begin{pgfonlayer}{nodelayer}
		\node [style=Z dot] (1) at (-1, 0) {};
		\node [style=Z dot] (2) at (-1, 1) {};
		\node [style=Z dot] (3) at (-1, 2) {};
		\node [style=X dot] (4) at (0, 1.5) {};
		\node [style=Z phase dot] (5) at (1.25, 1.5) {$\alpha$};

		\node [style=none] (6) at (3,1) {$=$};

		\node [style= Z dot] (7) at (5,2) {};
		\node [style= X dot] (8) at (5,1) {};
		\node [style= Z dot] (9) at (6,1) {};
		\node [style= X dot] (10) at (6,0) {};
		\node [style= Z phase dot] (11) at (7,0) {$\alpha$};
		\node [style= Z dot] (12) at (8,1) {};
		\node [style= X dot] (13) at (8,0) {};
		\node [style= Z dot] (14) at (9,2) {};
		\node [style= X dot] (15) at (9,1) {};

		\node [style= none] (16) at (10, 0) {};
		\node [style= none] (17) at (10, 1) {};
		\node [style= none] (18) at (10, 2) {};

		\node [style= none] (19) at (4, 0) {};
		\node [style= none] (20) at (4, 1) {};
		\node [style= none] (21) at (4, 2) {};

		\node [style= none] (22) at (-2, 0) {};
		\node [style= none] (23) at (-2, 1) {};
		\node [style= none] (24) at (-2, 2) {};

		\node [style= none] (25) at (2, 0) {};
		\node [style= none] (26) at (2, 1) {};
		\node [style= none] (27) at (2, 2) {};

	\end{pgfonlayer}
	\begin{pgfonlayer}{edgelayer}

		\draw  (16.center) to (19.center);
		\draw  (17.center) to (20.center);
		\draw  (18.center) to (21.center);

		\draw  (22.center) to (25.center);
		\draw  (23.center) to (26.center);
		\draw  (24.center) to (27.center);

		\draw [in=90, out=-5, looseness=0.75] (1) to (4.center);
		\draw [in=180, out=0] (2) to (4.center);
		\draw [in=-90, out=5, looseness=0.75] (3) to (4.center);
		\draw [in=0, out=0] (4) to (5.center);

		\draw [in=90, out=-90] (7) to (8.center);
		\draw [in=0, out=0] (8) to (9.center);

		\draw [in=90, out=-90] (9) to (10.center);
		\draw [in=0, out=0] (10) to (11.center);

		\draw [in=0, out=0] (11) to (13.center);
		\draw [out=90, in=-90] (13) to (12.center);

		\draw [out=90, in=-90] (15) to (14.center);
		\draw [in=0, out=0] (12) to (15.center);

	\end{pgfonlayer}
\end{tikzpicture}
    %}
  \end{center}
  \caption{Example of a phase gadget and its decomposition into a
  CNOT ladder.}
  \label{figphasegadget}
\end{figure}

\subsection{Rewriting Rules}
\label{sec:zxrules}
\begin{figure*}[t]
  \resizebox{\textwidth}{!}{
    \begin{tabular}{cccc}
      \emph{ID-removal} & \emph{Fusion} &   \emph{Bialgebra}\\
      \begin{tikzpicture}
	\begin{pgfonlayer}{nodelayer}
		\node [style=none] (31) at (-12, 4.75) {$\overset{(i1)}{=}$};
		\node [style=none] (32) at (-12, 3.25) {$\overset{(i2)}{=}$};

		\node [style=none] (36) at (-11, 2.25) {};
		\node [style=hadamard] (35) at (-11.75, 2.25) {};
		\node [style=hadamard] (41) at (-12.25, 2.25) {};
		\node [style=none] (37) at (-13, 2.25) {};

		\node [style=none] (38) at (-11, 5.75) {};
		\node [style=Z dot] (39) at (-12, 5.75) {};
		\node [style=none] (40) at (-13, 5.75) {};

		\node [style=none] (33) at (-11, 4) {};
		\node [style=none] (34) at (-13, 4) {};
	\end{pgfonlayer}
	\begin{pgfonlayer}{edgelayer}
		\draw (36.center) to (35);
		\draw (38.center) to (39);
		\draw (39) to (40.center);
		\draw (33.center) to (34.center);
		\draw (35) to (41);
		\draw (41) to (37.center);
	\end{pgfonlayer}
\end{tikzpicture} & \begin{tikzpicture}
	\begin{pgfonlayer}{nodelayer}
		\node [style=none] (11) at (-10.75, 3.5) {$\overset{(f)}{=}$};
		\node [style=Z phase dot] (12) at (-12, 3.5) {$\beta$};
		\node [style=none] (13) at (-11, 2.5) {};
		\node [style=none] (14) at (-12.75, 2.5) {};
		\node [style=Z phase dot] (15) at (-13.25, 4.25) {$\alpha$};
		\node [style=none] (16) at (-12.5, 5) {};
		\node [style=none] (17) at (-14, 5) {};
		\node [style=none] (18) at (-14, 2.5) {};
		\node [style=Z phase dot] (19) at (-9, 3.75) {$\alpha+\beta$};
		\node [style=none] (20) at (-10, 5) {};
		\node [style=none] (21) at (-10, 2.5) {};
		\node [style=none] (22) at (-8, 2.5) {};
		\node [style=none] (23) at (-8, 5) {};
		\node [style=none] (24) at (-13, 3.5) {\small$\ldots$};
		\node [style=none] (25) at (-13.25, 5) {\small$\ldots$};
		\node [style=none] (26) at (-9, 4.75) {\small$\ldots$};
		\node [style=none] (27) at (-9, 3) {\small$\ldots$};
		\node [style=none] (28) at (-12, 2.5) {\small$\ldots$};
		\node [style=none] (29) at (-11, 5) {};
		\node [style=none] (30) at (-12, 4.25) {\small$\ldots$};
	\end{pgfonlayer}
	\begin{pgfonlayer}{edgelayer}

		\draw [in=90, out=-5, looseness=0.75] (12) to (13.center);
		\draw [in=90, out=-175, looseness=0.75] (12) to (14.center);

		\draw [in=185, out=90, looseness=0.75] (18.center) to (15);

		\draw [in=-90, out=175, looseness=0.75] (15) to (17.center);
		\draw [in=-90, out=5, looseness=0.75] (15) to (16.center);

		\draw [in=150, out=-15] (15) to (12);

		\draw [in=90, out=-5, looseness=0.75] (19) to (22.center);
		\draw [in=90, out=-175, looseness=0.75] (19) to (21.center);
		\draw [in=-90, out=5, looseness=0.75] (19) to (23.center);
		\draw [in=-90, out=175, looseness=0.75] (19) to (20.center);

		\draw [in=-90, out=5] (12) to (29.center);
	\end{pgfonlayer}
\end{tikzpicture} &  \begin{tikzpicture}
	\begin{pgfonlayer}{nodelayer}
		\node [style=none] (42) at (-11, 3.5) {$\overset{(b)}{=}$};
		\node [style=none] (43) at (-11.25, 4.5) {};
		\node [style=none] (44) at (-11.25, 2.5) {};
		\node [style=Z dot] (45) at (-12.25, 3.5) {};
		\node [style=X dot] (46) at (-13, 3.5) {};
		\node [style=none] (47) at (-14, 4.5) {};
		\node [style=none] (48) at (-14, 2.5) {};
		\node [style=none] (49) at (-10.25, 4.5) {};
		\node [style=none] (50) at (-10.25, 2.5) {};
		\node [style=Z dot] (51) at (-9.75, 4.5) {};
		\node [style=Z dot] (52) at (-9.75, 2.5) {};
		\node [style=X dot] (53) at (-8.5, 2.5) {};
		\node [style=X dot] (54) at (-8.5, 4.5) {};
		\node [style=none] (55) at (-8, 4.5) {};
		\node [style=none] (56) at (-8, 2.5) {};
	\end{pgfonlayer}
	\begin{pgfonlayer}{edgelayer}
		\draw [in=95, out=0] (47.center) to (46);
		\draw [in=0, out=-95] (46) to (48.center);
		\draw (45) to (46);
		\draw [in=180, out=85] (45) to (43.center);
		\draw [in=180, out=-85] (45) to (44.center);
		\draw (50.center) to (52);
		\draw (52) to (54);
		\draw (49.center) to (51);
		\draw (51) to (53);
		\draw (53) to (56.center);
		\draw (54) to (55.center);
		\draw (51) to (54);
		\draw (52) to (53);
	\end{pgfonlayer}
\end{tikzpicture}\\[-3em]
        \emph{Copy} &  \emph{Pi-Copy}  & \emph{Color change} \\
      \begin{tikzpicture}
	\begin{pgfonlayer}{nodelayer}
		\node [style=none] (0) at (-11, 2) {$\overset{(c)}{=}$};
		\node [style=none] (1) at (-12, 2.75) {};
		\node [style=none] (2) at (-12, 2) {\vdots};
		\node [style=none] (3) at (-12, 0.75) {};
		\node [style=Z phase dot] (4) at (-13, 1.75) {$\alpha$};
		\node [style=X dot] (5) at (-14, 1.75) {};
		\node [style=none] (6) at (-9.25, 2.75) {};
		\node [style=none] (7) at (-10, 2) {\vdots};
		\node [style=none] (8) at (-9.25, 0.75) {};
		\node [style=X dot] (9) at (-10, 2.75) {};
		\node [style=X dot] (10) at (-10, 0.75) {};
	\end{pgfonlayer}
	\begin{pgfonlayer}{edgelayer}
		\draw (5) to (4);
		\draw [in=-180, out=85] (4) to (1.center);
		\draw [in=-180, out=-85] (4) to (3.center);
		\draw (9) to (6.center);
		\draw (10) to (8.center);
	\end{pgfonlayer}
\end{tikzpicture} & \begin{tikzpicture}
	\begin{pgfonlayer}{nodelayer}
		\node [style=none] (57) at (-10.75, 2) {$\overset{(\pi)}{=}$};
		\node [style=none] (58) at (-11.5, 3) {};
		\node [style=none] (59) at (-11.75, 2.25) {\vdots};
		\node [style=none] (60) at (-11.5, 1) {};
		\node [style=Z phase dot] (61) at (-12.75, 2) {$\alpha$};
		\node [style=X phase dot] (62) at (-13.75, 2) {$\pi$};
		\node [style=none] (63) at (-14.5, 2) {};
		\node [style=none] (64) at (-10, 2) {};
		\node [style=Z phase dot] (65) at (-9.25, 2) {$-\alpha$};
		\node [style=none] (66) at (-7.625, 3) {};
		\node [style=none] (67) at (-8.25, 2.25) {\vdots};
		\node [style=none] (68) at (-7.625, 1) {};
		\node [style=X phase dot] (69) at (-8.25, 3) {$\pi$};
		\node [style=X phase dot] (70) at (-8.25, 1) {$\pi$};
	\end{pgfonlayer}
	\begin{pgfonlayer}{edgelayer}
		\draw (63.center) to (62);
		\draw (62) to (61);
		\draw [in=-180, out=85] (61) to (58.center);
		\draw [in=-180, out=-85] (61) to (60.center);
		\draw (64.center) to (65);
		\draw [in=-180, out=85] (65) to (69);
		\draw (69) to (66.center);
		\draw [in=180, out=-85] (65) to (70);
		\draw (70) to (68.center);
	\end{pgfonlayer}
\end{tikzpicture} & \begin{tikzpicture}
	\begin{pgfonlayer}{nodelayer}
		\node [style=X phase dot] (71) at (-13.75, 1.75) {$\alpha$};
		\node [style=none] (76) at (-12.75, 2.0) {$\vdots$};
		\node [style=none] (77) at (-14.75, 2.0) {$\vdots$};
		\node [style=none] (78) at (-12.25, 1.75) {$\overset{(h)}{=}$};

		\node [style=none] (72) at (-12.75, 2.75) {};
		\node [style=none] (73) at (-14.75, 2.75) {};
		\node [style=none] (74) at (-14.75, 0.75) {};
		\node [style=none] (75) at (-12.75, 0.75) {};

		\node [style=none] (80) at (-11.75, 2.75) {};
		\node [style=none] (83) at (-11.75, 0.75) {};
		\node [style=none] (81) at (-9.75, 2.75) {};
		\node [style=none] (82) at (-9.75, 0.75) {};

		\node [style=Z phase dot] (79) at (-10.75, 1.75) {$\alpha$};
		\node [style=none] (84) at (-11.75, 2) {$\vdots$};
		\node [style=none] (85) at (-9.75, 2) {$\vdots$};

		\node [style=hadamard] (86) at (-11.75, 2.75) {};
		\node [style=hadamard] (87) at (-9.75, 0.75) {};
		\node [style=hadamard] (88) at (-9.75, 2.75) {};
		\node [style=hadamard] (89) at (-11.75, 0.75) {};

		\node [style=none] (90) at (-12.125, 2.75) {};
		\node [style=none] (91) at (-9.375, 0.75) {};
		\node [style=none] (92) at (-9.375, 2.75) {};
		\node [style=none] (93) at (-12.125, 0.75) {};

		\node [style=none] (10) at (-15.125, 2.75) {};
		\node [style=none] (11) at (-12.375, 0.75) {};
		\node [style=none] (12) at (-12.375, 2.75) {};
		\node [style=none] (13) at (-15.125, 0.75) {};
	\end{pgfonlayer}
	\begin{pgfonlayer}{edgelayer}
		\draw [in=180, out=85, looseness=0.75] (71) to (72.center);
		\draw [in=0, out=95] (71) to (73.center);
		\draw [in=0, out=-95, looseness=0.75] (71) to (74.center);
		\draw [in=180, out=-85, looseness=0.75] (71) to (75.center);

		\draw [in=0, out=95, looseness=0.75] (79) to (80.center);
		\draw [in=0, out=-95, looseness=0.75] (79) to (83.center);
		\draw [in=180, out=85, looseness=0.75] (79) to (81.center);
		\draw [in=180, out=-85, looseness=0.75] (79) to (82.center);

		\draw (86) to (90.center);
		\draw (87) to (91.center);
		\draw (88) to (92.center);
		\draw (89) to (93.center);

		\draw (75.center) to (11.center);
		\draw (72.center) to (12.center);
		\draw (73.center) to (10.center);
		\draw (74.center) to (13.center);
	\end{pgfonlayer}
\end{tikzpicture} \\
    \end{tabular}
  }
    \caption{The basic rewriting rules of the scalar-free $\frac{\pi}{2}$ ZX-calculus.}
  \label{fig:rules}
\end{figure*}
%\begin{itemize}
%    \item \xxx{Better example where more (preferable all) rules can be applied}
%\end{itemize}

This section introduces the basic rewriting rules of the ZX-calculus that are outlined in Figure~\ref{fig:rules}~\cite{coeckePicturingQuantumProcesses2017,staudacherReducing2QuBitGate2023}.

A rewriting rule transforms a ZX-diagram while preserving its underlying semantics under the following definition:
\begin{definition}
  \label{def:rewriting rule}
  %  Let $\mathbf{ZX}$ be the set of all labeled open graphs as defined in
  %  Definition~\ref{def:opengraph}.
    Let $\mathbf{LM}$ be the set of linear maps among qubits, and let $\mathbf{ZX}$ be the set of ZX-diagrams.
  The function $\gamma : \mathbf{ZX} \to \mathbf{LM}$ is a function that maps a ZX-diagram to its linear map between qubits.
  Two ZX-diagrams that represent the same linear map between qubits $g,h \in \mathbf{ZX}, \gamma(g)=\gamma (h)$ have the same semantics.
  A rewriting rule is a function $r: \mathbf{ZX} \to \mathbf{ZX}$ that transforms a ZX-diagram while preserving its underlying semantics, such that $g\in\mathbf{ZX}, \gamma (g) = (\gamma \circ r)(g)$.
\end{definition}

\begin{figure*}[t!]
  \centering
  %\begin{center}
  \resizebox{\textwidth}{!}{
    \begin{tikzpicture}
	\begin{pgfonlayer}{nodelayer}
        \node (100) at (0,0) {
                
            $\Qcircuit @C=0.25em @R=.25em @!R {
                & \gate{T}  & \ctrl{1}  & \gate{T}  & \qw & \targ    & \qw \\
                & \qw       & \targ     & \gate{T}  & \qw & \ctrl{-1} & \qw}$

        };
        
        % input zx diagram
        \node [style=none] (0) at ([shift={(3,0)}]0.00, 0.60) {};
        \node [style=none] (1) at ([shift={(3,0)}]0.00, -0.60) {};
        \node [style=Z phase dot] (2) at ([shift={(3,0)}]1.00, 0.60) {$\frac{\pi}{4}$};
        \node [style=X dot] (3) at ([shift={(3,0)}]2.00, -0.60) {};
        \node [style=Z dot] (4) at ([shift={(3,0)}]2.00, 0.60) {};
        \node [style=Z phase dot] (5) at ([shift={(3,0)}]3.00, 0.60) {$\frac{\pi}{4}$};
        \node [style=Z phase dot] (6) at ([shift={(3,0)}]3.00, -0.60) {$\frac{\pi}{4}$};
        \node [style=X dot] (7) at ([shift={(3,0)}]4.00, 0.60) {};
        \node [style=Z dot] (8) at ([shift={(3,0)}]4.00, -0.60) {};
        \node [style=none] (9) at ([shift={(3,0)}]5.00, 0.60) {};
        \node [style=none] (10) at ([shift={(3,0)}]5.00, -0.60) {};
        \node [style=none] (15) at (2.75,-0.3) {$=$};

        %  spider fusion
        \node [style=none] (110) at ([shift={(9,0)}]0.00, 0.60) {};
        \node [style=none] (11) at ([shift={(9,0)}]0.00, -0.60) {};
        \node [style=X dot] (13) at ([shift={(9,0)}]1.50, -0.60) {};
        \node [style=Z phase dot] (14) at ([shift={(9,0)}]1.50, 0.60) {$\frac{\pi}{2}$};
        \node [style=X dot] (17) at ([shift={(9,0)}]3.50, 0.60) {};
        \node [style=Z phase dot] (18) at ([shift={(9,0)}]3.50, -0.60) {$\frac{\pi}{4}$};
        \node [style=none] (19) at ([shift={(9,0)}]5.00, 0.60) {};
        \node [style=none] (111) at ([shift={(9,0)}]5.00, -0.60) {};
        \node [style=none] (15) at ([shift={(6,0)}]2.5,0) {$\overset{(f)}{=}$};

        % input colour change
        \node [style=none] (210) at ([shift={(15,0)}]0.00, 0.60) {};
        \node [style=none] (21) at ([shift={(15,0)}]0.00, -0.60) {};
        \node [style=Z dot] (23) at ([shift={(15,0)}]1.50, -0.60) {};
        \node [style=Z phase dot] (24) at ([shift={(15,0)}]1.50, 0.60) {$\frac{\pi}{2}$};
        \node [style=Z dot] (27) at ([shift={(15,0)}]3.50, 0.60) {};
        \node [style=Z phase dot] (28) at ([shift={(15,0)}]3.50, -0.60) {$\frac{\pi}{4}$};
        \node [style=none] (29) at ([shift={(15,0)}]5.00, 0.60) {};
        \node [style=none] (211) at ([shift={(15,0)}]5.00, -0.60) {};
        \node [style=hadamard] at ([shift={(15,0)}] 2.5, 0.60) {};
        \node [style=hadamard] at ([shift={(15,0)}] 1.5, 0.00) {};
        \node [style=hadamard] at ([shift={(15,0)}] 3.5, 0.00) {};
        \node [style=hadamard] at ([shift={(15,0)}] 2.5, -0.60) {};
        \node [style=hadamard] at ([shift={(15,0)}] 0.5, -0.60) {};
        \node [style=hadamard] at ([shift={(15,0)}] 4.5, 0.60) {};
        \node [style=none] (15) at ([shift={(12,0)}]2.5,0) {$\overset{(h)}{=}$};

	\end{pgfonlayer}
	\begin{pgfonlayer}{edgelayer}
        % input zx diagram
        \draw (0) to (2);
        \draw (1) to (3);
        \draw (2) to (4);
        \draw (3) to (4);
        \draw (3) to (6);
        \draw (4) to (5);
        \draw (5) to (7);
        \draw (6) to (8);
        \draw (7) to (8);
        \draw (7) to (9);
        \draw (8) to (10);

        % spider fusion
        \draw (11) to (13);
        \draw (110) to (14);
        \draw (13) to (14);
        \draw (13) to (18);
        \draw (14) to (17);
        \draw (17) to (18);
        \draw (17) to (19);
        \draw (18) to (111);

        % colour change zx diagram
        \draw (21) to (23);
        \draw (210) to (24);
        \draw (23) to (24);
        \draw (23) to (28);
        \draw (24) to (27);
        \draw (27) to (28);
        \draw (27) to (29);
        \draw (28) to (211);
	\end{pgfonlayer}
\end{tikzpicture}
  }
  \caption{Successive applications of rewriting rules to a simple ZX
  diagram (to be read from left to right).} % and top to bottom).}
  \label{fig:running-example}
  %\end{center}
\end{figure*}
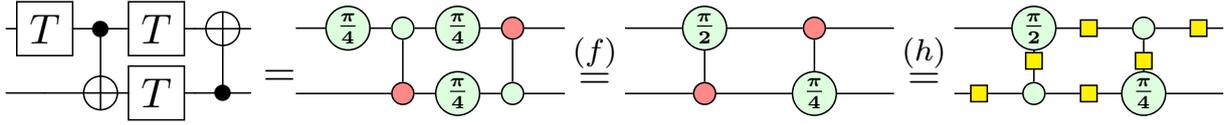

All rules remain valid under color inversion.
We give an example of the successive application of some rewriting rules on a simple ZX-diagram in Figure~\ref{fig:running-example}.

\paragraph*{\emph{Spider fusion (f)}}
Connected spiders of the same color fuse through modulo-$2\pi$ addition of their phases. 
The reverse unfusing operation is always possible, because connecting additional spiders with a phase of $\alpha=0$ will not change the modulo-$2\pi$ addition.
As a consequence, infinite spiders can be unfused.
Figure~\ref{fig:running-example} highlights the fusion of two green non-phase-carrying spiders with their neighboring phase-carrying spiders.

%\paragraph*{\emph{Local complementation (lc)}}
%The local complementation rule~\cite{kotzigEulerianLinesFinite1968}
%originates from graph theory. For all directly connected spiders of a
%target spider, local complementation connects previously unconnected
%spiders and disconnects previously connected spiders.
%The local complementation of the highlighted red spider in the bottom
%qubit row is illustrated in
%Figure~\ref{fig:running-example}. The two green spiders connected to
%the highlighted red spider connect via local complementation.
%Performing a second local complementation on the same red spider
%would disconnect the two green phase-carrying spiders again.
%Pivoting describes a series of local complementations.

\paragraph*{\emph{Color change (h)}}
Adding Hadamard generators to each input and output inverts the color of a spider. 
In Figure~\ref{fig:running-example}, all red spiders turn green with the addition of Hadamard generators.

\paragraph*{\emph{Identity removal (i1, i2)}}
Phase-less spiders  with a single input and output wire ($n=m=1$) implement the identity linear map. 
Therefore, they act like wires and the phase-less spider can be removed.
Similarly, two directly connected Hadamard generators cancel each other out and act like a wire.
%Furthermore, identity rules are used to convert a ZX-diagram to be graph-like~(Definition~\ref{def:graph-like}) by ensuring that spiders always connect with each other through Hadamard generators and by adding potentially missing spiders at the input and output.

%\begin{itemize}
%    \item \xxx{I want to have a better explanationton for Hopf and Bialgebra}
%    \item \xxx{explain what commutation relation between COPY and XOR algebra means}
%\end{itemize}

%\paragraph*{\emph{Hopf (ho)}}
%The Hopf rule originates from the \emph{COPY} and \emph{XOR}
%algebras that form a Hopf algebra.
%If multiple wires connect two opposite colored spiders, the
%additional wires can be removed pair by pair.

\paragraph*{\emph{Bialgebra (b)}}
The bialgebra rule originates from the commutation relation between the \emph{COPY} and \emph{XOR} algebra.
This rule permits connected spiders of opposite colors to move through each other at the cost of potentially adding spiders.

\paragraph*{\emph{Copy ($\pi$)}}
$\pi$ copying moves an input spider that carries the phase $\alpha=\pi$ through an opposite colored spider to all connected wires while multiplying the phase by $-1$.

\paragraph*{\emph{State copy (c)}}
State copy is a special case of the copy rule when the input spider does not have any input wire ($n=0$) and the phase is a multiple of $\pi$.
It copies the computational basis through an opposite-colored spider and places the copied spider on every outgoing wire.
The opposite-colored spider vanishes.

%\paragraph*{\emph{Neighbor unfusion}}
%\citet{staudacherReducing2QuBitGate2023} introduce the neighbor unfusion rule, a complement to the spider fusion rule, which enables further optimizations.
%Subsequent research by~\citet{mcelvanneyFlowpreservingZXcalculusRewrite2023} proves that neighbor unfusion also preserves general flow. This important result allows for the application of the unfusion rule without requiring repeated computationally expensive general flow verifications, thus accelerating the optimization workflow.

\subsubsection{ZX-based quantum circuit Optimization}
At the heart of ZX-based quantum circuit optimization is the successive application of rewriting rules.
%To simplify ZX-diagrams using basic rewriting rules, the popular PyZX framework assumes that the ZX-diagrams are graph-like~(Definition~\ref{def:graph-like})~\cite{kissingerPyZXLargeScale2020}.
Figure~\ref{fig:running-example} shows an example rewriting rule sequence of that  applies the fusion (f) and color change (h) rules and results in a graph-like ZX-diagram~(Definition~\ref{def:graph-like}). 

\paragraph{Optimization Pipeline} Figure~\ref{fig:optimization} shows an idealized three-step optimization pipeline for ZX-based quantum circuit optimization.
\begin{enumerate}
  \item \emph{Convert:} The quantum circuit is converted to an equivalent ZX-diagram. ZX-calculus is universal; hence every quantum circuit can be converted.
  \item \emph{Search:} ZX-diagram optimization for different target metric(s) by applying semantic preserving rewriting     rules~(Figure~\ref{fig:rules})
  \item \emph{Extraction:} Circuit extraction~(Section~\ref{chap1:sec:circuit-extraction-problem}) translates the ZX-diagram into an equivalent quantum circuit.
    %Although only diagrams that preserve general flow or causal flow are extractable in polynomial-time otherwise it is a combinatorial optimization problem that is \#P-hard.
\end{enumerate}
The two entry points to optimize ZX-diagrams are the search stage and the extraction stage.
The search stage primarily targets ZX-diagram metrics that directly translate or approximate quantum circuit metrics (e.g., Non-Clifford spiders and Hadamard wires).
The circuit extraction stage targets quantum circuit metrics (e.g., two-qubit gate count and circuit depth) that are possibly architecture-aware.

\subsection{The Circuit Extraction Problem}
\label{sec:circuit-extraction-problem}

The conversion of ZX-diagrams into quantum circuits necessitates computationally expensive circuit extraction algorithms.
A ZX-diagram does not have one but many valid quantum circuit representations.
As a consequence, the circuit characteristics and optimization results are heavily reliant on the chosen extraction method.
This section introduces the principles of the circuit extraction algorithm proposed by Duncan et al.~\cite{duncanGraphtheoreticSimplificationQuantum2020}.
Its extension by Backens et al.~\cite{backensThereBackAgain2020} is included as the standard algorithm in many works, as it allows for the extraction of phase gadgets.

In principle, circuit extraction is \#P-hard and poses an optimization problem with an upper bound of $\mathrm{NP}^{\mathrm{NP}^{\#\mathrm{P}}}$~\cite{debeaudrapCircuitExtractionZXdiagrams2022,mitosek2024constructing}.
Therefore, it is computationally beneficial to reduce the need of circuit extraction as much as possible.
Nevertheless, there are polynomial-time circuit extraction algorithms for ZX-diagrams in the MBQC form that maintain a graph-theoretic property known as general flow or the stricter causal flow~\cite{backensThereBackAgain2020}.

These algorithms allow for the extraction of larger ZX-diagrams, but do so at the cost of introducing additional two-qubit gates to preserve the connectivity of spiders.
As a result, these extraction algorithms can lead to circuits with an increased gate count and depth compared to the source circuit.

\subsubsection{Measurement-based Quantum Computing}
Measurement-based quantum computing (MBQC), also known as one-way quantum computing, is an alternative to the quantum circuit model~\cite{raussendorfOneWayQuantumComputer2001,briegelMeasurementbasedQuantumComputation2009}.

In this model, the computation is performed in two steps: (i) the preparation of a highly entangled resource state, often a graph state, (ii) followed by sequential single-qubit measurements known as measurement patterns.
Quantum measurements are inherently non-deterministic, and the measurement outcomes might introduce unwanted behavior in the remaining unmeasured computation.
To ensure deterministic behavior, future qubits are only corrected based on previous measurement outcomes without affecting already measured qubits.
The measurements itself can be carried out in different measurement planes (XZ-, XY-, YZ-planes).
Determinism in the MBQC model requires sequence of measurement patterns where all qubits are measured, while only correcting future qubits at each measurement.
For a detailed introduction into MBQC, we refer to the paper by Wei~\cite{wei2021mbqc}.
A major application of MBQC is photonic quantum computing~\cite{couteau2025photonic}.

Following the work of Backens et al.~\cite{backensThereBackAgain2020},  ZX-diagrams can be interpreted as graph states (Definition~\ref{def:graph-state}) in the framework of MBQC.
\begin{definition} 
   \label{def:graph-state}
    \emph{Graph state diagram~\cite{backensThereBackAgain2020}}
    \begin{enumerate}
        \item All spiders are Z-spiders.
        \item Only Hadamard edges between spiders.
        \item Each spider has a single incident output wire.
    \end{enumerate}
\end{definition}
A ZX-diagram is in the \emph{MBQC form} if it can be interpreted as a graph state \emph{and} each spider is allowed to be connected to an input and a measurement effect instead of the output.
If a ZX-diagram is in the MBQC form and the placement of single-qubit Clifford gates on the inputs and outputs is allowed, it is named \emph{MBQC+LC (local Clifford) form}. 

ZX-diagrams in the MBQC form are labelled open graphs (Definition~\ref{def:opengraph})
\begin{definition}
  \label{def:opengraph}
  \emph{Open graph~\cite{dixon2013opengraph}}
  Let $G(\mathbf{V},\mathbf{E},\mathbf{I},  \mathbf{O},\lambda)$ be a finite undirected graph formed by the set of vertices $\mathbf{V}$, edges $\mathbf{E}$, inputs $\mathbf{I}\subseteq\mathbf{V}$ and outputs $\mathbf{O}\subseteq\mathbf{V}$.
  Let $deg: \mathbf{V}\rightarrow \mathbb{N}$ be a function that retrieves the degree, the number of connected edges, of a vertex.
  The sets of inputs $\mathbf{I}$ and outputs $\mathbf{O}$ only consist of single degree vertices, such that $v\in\mathbf{I}\cup\mathbf{O},\text{deg} (v)=1$.
  Together, the set of input and output vertices $\mathbf{I}\cup\mathbf{O}$ form the boundary of $G$.
    Let $\mathbf{\tilde{O}}=\mathbf{\mathbf{V} \setminus (\mathbf{I}\cup\mathbf{O})}$ be the set of interior vertices.
  Let $\lambda$ be a labeling function.
    The function $\lambda: \mathbf{\tilde{O}}\to \{XY, XZ, YZ\}$ assigns a measurement plane.
  We denote by $\mathbf{ZX}$ the set of all labeled open graphs.
\end{definition}

A ZX-diagram is deterministic when its underlying open graph admits a sequence of valid measurement patterns until all qubits are measured. 

ZX-diagrams do not have a temporal or spatial order of spiders beyond its connectivity; hence no partial order can be inferred trivially.
In contrast, the quantum circuit model places gates that act son specific qubits.
This representation implicitly defines partial order on the gates since they are arranged by time.

When interpreting a ZX-diagram in the MBQC form, a sequence of deterministic measurement patterns induces a partial execution order on the vertices.
In general, multiple partial orders can be consistent with a ZX-diagram's underlying open graph.

The presence of causal, general or Pauli flow inside a ZX-diagram's open graph are sufficient conditions for deterministic behavior~\cite{duncanGraphtheoreticSimplificationQuantum2020,backensThereBackAgain2020,simmons2021Measurement}.
For MBQC diagrams that admit one of these flows, polynomial-time circuit extraction algorithms exist.
Circuit extraction is not a mere translation between representations, instead it asserts deterministic behavior and induces a valid execution order.

A graph-like ZX-diagram~(Definition~\ref{def:graph-like}) corresponds to a ZX-diagram in the MBQC form where all measurements are performed in the XY-plane. 
For graph-like ZX-diagrams, the existence of \emph{causal flow} suffices to guarantee a deterministic measurement pattern~\cite{danos2006causal}.
In the general case, ZX-diagrams in the MBQC form contain measurements in the XZ-, XY-, YZ- planes.
The existence of \emph{general flow}~\cite{browne2007general} is a sufficient condition for deterministic measurement patterns in the general case.

\emph{For the rest of this review paper, we assume that ZX-diagrams are in the MBQC form.}

\subsubsection{Causal and General Flow}
\label{chap1:sec:flow}
\begin{figure*}[ht!]
  \begin{subfigure}[t]{0.3\textwidth}
    \begin{tikzpicture}
    \begin{pgfonlayer}{nodelayer}
        \node [style=none] (1) at (2.00, -0.00) {};
        \node [style=none] (2) at (2.00, -2.00) {};
        \node [style=Z dot, label=above left:{1}] (4) at (4.00, -0.00) {};
        \node [style=Z dot, label=above right:{4}] (5) at (6.00, -0.00) {};
        \node [style=Z dot, , label=above right:{5}] (6) at (6.00, -2.00) {};
        \node [style=Z dot, label=above left:{2}] (7) at (4.00, -2.00) {};
        \node [style=none] (10) at (8.00, -0.00) {};
        \node [style=none] (11) at (8.00, -2.00) {};
        \node [style=Z phase dot] (13) at (4.00, -1.00) {$\beta$};
        \node [style=Z phase dot] (15) at (4.00, 1.00) {$\alpha$};
        \node [style=none] (3) at (2.00, -4.00) {};
        \node [style=Z dot, label=above left:{3}] (8) at (4.00, -4.00) {};
        \node [style=Z dot, label=above right:{6}] (9) at (6.00, -4.00) {};
        \node [style=none] (12) at (8.00, -4.00) {};
        \node [style=Z phase dot] (14) at (4.00, -3.00) {$\gamma$};
    \end{pgfonlayer}
    \begin{pgfonlayer}{edgelayer}
        \draw (1) to (4);
        \draw (2) to (7);
        \draw (3) to (8);
        \draw (4) to (15);
        \draw [style=hadamard edge] (4) to (5);
        \draw (5) to (10);
        \draw [style=hadamard edge] (5) to (7);
        \draw (6) to (11);
        \draw [style=hadamard edge] (6) to (7);
        \draw [style=hadamard edge] (6) to (8);
        \draw (7) to (13);
        \draw (8) to (14);
        \draw [style=hadamard edge] (8) to (9);
        \draw (9) to (12);
    \end{pgfonlayer}
\end{tikzpicture}
    \caption{ZX-Diagram with causal flow.}
    \label{fig:causal-flow-example}
  \end{subfigure}
  ~
  \begin{subfigure}[t]{0.3\textwidth}
    \begin{tikzpicture}
    \begin{pgfonlayer}{nodelayer}
        \node [style=none] (1) at (2.00, -0.00) {};
        \node [style=none] (2) at (2.00, -2.00) {};
        \node [style=Z dot, label=above left:{1}] (4) at (4.00, -0.00) {};
        \node [style=Z dot, label=above right:{4}] (5) at (6.00, -0.00) {};
        \node [style=Z dot, , label=above right:{5}] (6) at (6.00, -2.00) {};
        \node [style=Z dot, label=above left:{2}] (7) at (4.00, -2.00) {};
        \node [style=none] (10) at (8.00, -0.00) {};
        \node [style=none] (11) at (8.00, -2.00) {};
        \node [style=Z phase dot] (13) at (4.00, -1.00) {$\beta$};
        \node [style=Z phase dot] (15) at (4.00, 1.00) {$\alpha$};
        \node [style=none] (3) at (2.00, -4.00) {};
        \node [style=Z dot, label=above left:{3}] (8) at (4.00, -4.00) {};
        \node [style=Z dot, label=above right:{6}] (9) at (6.00, -4.00) {};
        \node [style=none] (12) at (8.00, -4.00) {};
        \node [style=Z phase dot] (14) at (4.00, -3.00) {$\gamma$};
    \end{pgfonlayer}
    \begin{pgfonlayer}{edgelayer}
        \draw (1) to (4);
        \draw (2) to (7);
        \draw (4) to (15);
        \draw [style=hadamard edge] (4) to (5);
        \draw [style=hadamard edge] (4) to (9);
        \draw (5) to (10);
        \draw [style=hadamard edge] (5) to (7);
        \draw (6) to (11);
        \draw [style=hadamard edge] (6) to (7);
        \draw [style=hadamard edge] (6) to (8);
        \draw (7) to (13);
        \draw [style=hadamard edge] (7) to (9);
        \draw (3) to (8);
        \draw (8) to (14);
        \draw [style=hadamard edge] (8) to (9);
        \draw (9) to (12);
    \end{pgfonlayer}
\end{tikzpicture}
    \caption{ZX-diagram with general flow.}
    \label{fig:general-flow-example}
  \end{subfigure}
  ~
  \begin{subfigure}[t]{0.3\textwidth}
    \begin{tikzpicture}
    \begin{pgfonlayer}{nodelayer}
        \node [style=none] (1) at (2.00, -0.00) {};
        \node [style=none] (2) at (2.00, -2.00) {};
        \node [style=Z dot, label= above left:{1}] (4) at (4.00, -0.00) {};
        \node [style=Z dot, label= above right:{4}] (5) at (6.00, -0.00) {};
        \node [style=Z dot, label= above right:{5}] (6) at (6.00, -2.00) {};
        \node [style=Z dot, label= above left:{2}] (7) at (4.00, -2.00) {};
        \node [style=none] (10) at (8.00, -0.00) {};
        \node [style=none] (11) at (8.00, -2.00) {};
        \node [style=Z phase dot] (13) at (4.00, -1.00) {$\beta$};
        \node [style=Z phase dot] (15) at (4.00, 1.00) {$\alpha$};
        \node [style=none] (12) at (8.00, -4.00) {};
        \node [style=Z phase dot] (14) at (4.00, -3.00) {$\gamma$};
        \node [style=Z dot, label=above left:{3}] (8) at (4.00, -4.00) {};
        \node [style=Z dot, label=above right:{6}] (9) at (6.00, -4.00) {};
        \node [style=none] (3) at (2.00, -4.00) {};
    \end{pgfonlayer}
    \begin{pgfonlayer}{edgelayer}
        \draw (1) to (4);
        \draw (2) to (7);
        \draw (3) to (8);
        \draw (4) to (15);
        \draw [style=hadamard edge] (4) to (5);
        \draw [style=hadamard edge] (4) to (6);
        \draw (5) to (10);
        \draw [style=hadamard edge] (5) to (7);
        \draw (6) to (11);
        \draw [style=hadamard edge] (6) to (8);
        \draw [style=hadamard edge] (6) to (7);
        \draw [style=hadamard edge] (7) to (9);
        \draw [style=hadamard edge] (8) to (9);
        \draw [style=hadamard edge] (8) to (5);
        \draw [style=hadamard edge] (4) to (9);
        \draw (7) to (13);
        \draw (9) to (12);
        \draw (8) to (14);
    \end{pgfonlayer}
\end{tikzpicture}
    \caption{ZX-diagram without flow.}
    \label{fig:no-flow-example}
  \end{subfigure}
  \caption{The different flow states of ZX-diagrams.}
\end{figure*}

%General and causal flow of
%A ZX-diagram $ZX$ admits causal or general flow if there is partial order that implements a deterministic measurement pattern
%
%Let $g: \mathbf{V\setminus O} \rightarrow \{\mathbf{S} \subseteq\mathcal{P}\left(\mathbf{V\setminus I}\right) \mid |\mathbf{S}|=1 \}$ be a mapping

\paragraph{Causal flow}
Causal flow is a sufficient condition for the existence of a deterministic measurement pattern for ZX-diagrams that are restricted to measurements in the XY-plane.
Deterministic measurement patterns that always correct a single unmeasured qubit admit causal flow.

\begin{definition}[Causal Flow~\cite{danos2006causal}]
  \label{def:causal-flow}
  Let $ZX$ be an open graph with the set of vertices $\mathbf{V}$, inputs $\mathbf{I}$, and outputs $\mathbf{O}$.
  The function $g: \mathbf{V\setminus O} \rightarrow \mathbf{V\setminus I}$ maps measured qubits to unmeasured correction qubits.
  The neighborhood $N(v)$ of a vertex $v\in\mathbf{V}$ is the set of all adjacent vertices.
  causal flow is a strict partial order $\prec$ on $\mathbf{V}$, such that $\forall v \in \mathbf{V\setminus O}$:
  \begin{enumerate}
    \item $v\prec g\left(v\right)$
    \item if $w\in N\bigl(g\left(v\right)\bigr) \implies \left(w=v\right) \lor \left(v\prec w\right)$
    \item $v \in N\bigl(g\left(v\right)\bigr)$
  \end{enumerate}
\end{definition}

Figure~\ref{fig:causal-flow-example} shows a ZX-diagram that admits causal flow.
Following the causal flow Definition~\ref{def:causal-flow}, suppose $v=1$ is the first qubit to be corrected with the correction vertex $g(1)=4$. This implies the partial order of $1 \prec 4$.
Let the $v=2$ be the second qubit to be corrected by the correction vertex $g(2)=5$, implying an ordering of $2\prec 5$.
The final qubit $v=3$ is corrected with correction vertex $g(3)=6$, implying an ordering of $3\prec 6$.
Combining all partial orders lead to a causal flow with a measurement order of $1\prec 2 \prec 3$.

\paragraph{General flow}
General flow is a necessary condition for the existence of a deterministic measurement pattern of general ZX-diagrams, including other measurement planes.
In contrast to causal flow, general flow allows for the correction of multiple unmeasured qubits as a result of a single measurement.
\begin{definition}[General Flow~\cite{browne2007general,danos2009extendedMeasurement}]
  \label{def:general-flow}
  Let $ZX$ be an open graph with the set of vertices $\mathbf{V}$, inputs $\mathbf{I}$, and outputs $\mathbf{O}$.
  The function $g: \mathbf{V\setminus O} \rightarrow \mathcal{P}(\mathbf{V\setminus I})$ maps measured qubits to a correction set of unmeasured qubits.
  Let $\lambda : \mathbf{V} \rightarrow \{XY, XZ, YZ \}$ be a function that assigns a measurement plane to each vertex.
  Let $\tilde{N}(\mathbf{S})$ be a function that maps the correction set $\mathbf{S}\subseteq \mathbf{V}$ to the set of vertices that have an odd number of neighbors in $S$, such $\tilde{N}(\mathbf{S})=\{\forall u\in \mathbf{V}, | N(u) \cap \mathbf{S}| = 1\mod{2} \}$ where $N(u)$ denotes the neighborhood of vertex $u$.
  general flow is a partial order $\prec$ on $\mathbf{V}$, such that $\forall v \in \mathbf{V\setminus O}$:
  \begin{enumerate}
    \item if $w \in g\left(v\right) \land v \neq w \implies v\prec w$
    \item if $w\in \tilde{N}\bigl(g\left(v\right)\bigr) \land v \neq w  \implies v\prec w$
    \item if $\lambda \left(v\right) = \text{XY} \implies v\not\in g\left(v\right) \land v\in \tilde{N}\bigl(g\left(v\right)\bigr)$
    \item if $\lambda \left(v\right) = \text{XZ} \implies v\in g\left(v\right) \land v\in \tilde{N}\bigl(g\left(v\right)\bigr)$
    \item if $\lambda \left(v\right) = \text{YZ} \implies v\in g\left(v\right) \land v\not\in \tilde{N}\bigl(g\left(v\right)\bigr)$
  \end{enumerate}
\end{definition}

The ZX-diagram visualized in Figure~\ref{fig:general-flow-example} admits general flow.
Enumeration over all correction sets and subsequent filtering by the conditions of the general flow (Definition~\ref{def:general-flow}), it can be shown that the ZX-diagram admits 25 different general flows.
In the following, one general flow is considered in more detail.
Let qubit $v=2$ be the first qubit to be corrected with the correction set $g(2)=\{4\}$.
The resulting odd neighborhood $\tilde{N}\bigl(g(2)\bigr)=\{1,2\}$ implies an ordering of $2\prec 1$.
Next, qubit $v=1$ is to be corrected. With the choice of $g(2)=\{4\}$, $g(1)$ can take two different non-contradictory values, namely $\{5,6\}$ and $\{4,5\}$.
Let's correct qubit $v=1$ with the correction set $g(1)=\{4,5\}$, such that the odd neighborhood takes the form of $\tilde{N}\bigl(g(1)\bigr)=\{1,3\}$, implying $1\prec 3$.
Consequently, the last remaining qubit $v=3$ can only be corrected by $g(3)=\{4,6\}$ without effecting previously measured qubits with the odd neighborhood $\tilde{N}\bigl(g(3)\bigr)=\{3\}$.
The transitive closure of the implied constraints lead to a general flow with a measurement order of $2\prec 1 \prec 3$.

\paragraph{No flow}
ZX-diagrams are non-deterministic in the absence of causal flow, general flow or Pauli flow~\cite{duncanGraphtheoreticSimplificationQuantum2020,backensThereBackAgain2020,simmons2021Measurement}.
Figure~\ref{fig:no-flow-example} shows a ZX-diagram that does not admit any flow.
Intuitively, this can be seen from the fact that all input qubits are connected to all outputs.
It is impossible to find a measuring pattern without correcting on previously measured qubits.

To illustrate the absence of flow, consider the following example.
Let $v=1$  be the first qubit to be corrected by the correction set $g(1)=\{4\}$ with the corresponding odd neighborhood $\tilde{N}\bigl(g(1) \bigr)= \{1,2,3\}$.
Following the general flow Definition~\ref{def:general-flow}, an ordering of $1\prec 2$ and $1\prec 3$ is implied.
Let $v=2$ be the second qubit to be corrected by the correction set $g(2)= {5}$, resulting in the odd neighborhood of $\tilde{N}\bigl(g(2) \bigr)= \{1,2,3\}$.
This neighborhood implies an ordering of $2\prec 1$ and $2 \prec 3$.
Both orderings are contradictory, as they demand $1\prec 2$ and $2\prec 1$.
Even when enumerating all possible combinations of correction sets, it is impossible to find an ordering that does not contradict the general flow Definition~\ref{def:general-flow}.
Therefore, the ZX-diagram visualized in Figure~\ref{fig:no-flow-example} is non-deterministic and not extractable.

%ZX-calculus allows one to translate between MBQC and the quantum circuit model if a deterministic measurement pattern exists~\cite{ross2010measurement}.
%Circuit extraction is only possible for ZX-diagrams where it is possible to assert a deterministic measurement pattern; hence causal flow or general flow are preserved.

\subsubsection{Circuit Extraction in Polynomial-Time}
\label{chap1:sec:circuit-extraction-problem}
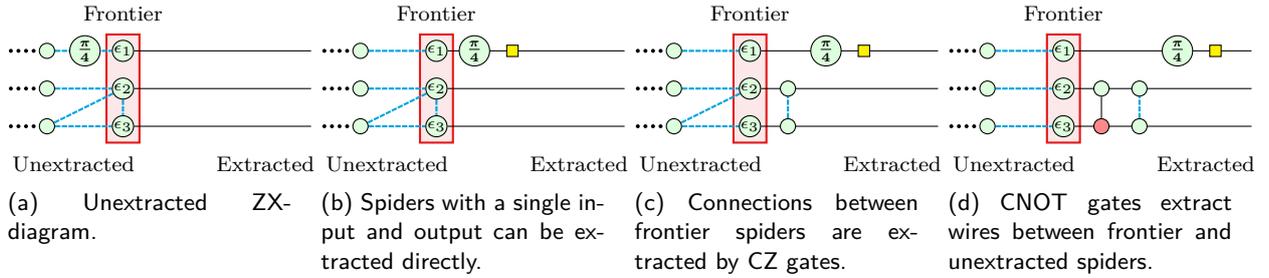
\begin{figure*}[ht!]
  \centering
  \begin{subfigure}[t]{0.48\textwidth}
    \centering
    \begin{tikzpicture}
	\begin{pgfonlayer}{nodelayer}
	    \node [style=Z dot] (0) at (0,0) {\tiny $\epsilon_{3}$};
	    \node [style=Z dot] (1) at (0,1) {\tiny $\epsilon_{2}$};
            \node [style=Z dot] (2) at (0,2) {\tiny $\epsilon_{1}$};
	    \node [style=Z phase dot] (3) at (-1,2) {$\frac{\pi}{4}$};

	    \node [style=none] (80) at (0,3) {{\scriptsize Frontier}};
	    \node [style=none] (81) at (-1.3,-1) {{\scriptsize Unextracted}};
	    \node [style=none] (82) at (3.75,-1) {{\scriptsize Extracted}};

	    \node [style=Z dot] (10) at (-2,0) {};
	    \node [style=Z dot] (11) at (-2,1) {};
	    \node [style=Z dot] (12) at (-2,2) {};

	    \node [style=none] (20) at (-3,0) {};
	    \node [style=none] (21) at (-3,1) {};
	    \node [style=none] (22) at (-3,2) {};

	    \node [style=none] (30) at (5,0) {};
	    \node [style=none] (31) at (5,1) {};
	    \node [style=none] (32) at (5,2) {};

        \begin{scope}[on background layer]
        \node[draw, color=luxembourg red, fill=luxembourg red!10, thick, inner sep=2pt, fit=(0) (1) (2)] (3) {};
        \end{scope}

	\end{pgfonlayer}
	\begin{pgfonlayer}{edgelayer}
	    \path[dotted pattern] (20) to (10);
	    \path[dotted pattern] (21) to (11);
	    \path[dotted pattern] (22) to (12);

	    \draw[style=hadamard edge] (10) to (0);
	    \draw[style=hadamard edge] (11) to (1);
	    \draw[style=hadamard edge] (12) to (2);

	    \draw[style=hadamard edge] (1) to (0);
	    \draw[style=hadamard edge] (1) to (10);

	    \draw (0) to (30);
	    \draw (1) to (31);
	    \draw (2) to (32);

	\end{pgfonlayer}
\end{tikzpicture}
    \caption{Unextracted ZX-diagram.}
    \label{a}
  \end{subfigure}
  ~
  \begin{subfigure}[t]{0.48\textwidth}
    \centering
    \begin{tikzpicture}
	\begin{pgfonlayer}{nodelayer}
	    \node [style=Z dot] (0) at (0,0) {\tiny $\epsilon_{3}$};
	    \node [style=Z dot] (1) at (0,1) {\tiny $\epsilon_{2}$};
	    \node [style=Z dot] (2) at (0,2) {\tiny $\epsilon_{1}$};
            \begin{scope}[on background layer]
                \node[draw, color=luxembourg red, fill=luxembourg red!10, thick, inner sep=2pt, fit=(0) (1) (2)] (3) {};
            \end{scope}
	    \node [style=hadamard] (4) at (2,2) {};
	    \node [style=Z phase dot] (5) at (1,2) {$\frac{\pi}{4}$};

	    \node [style=Z dot] (10) at (-2,0) {};
	    \node [style=Z dot] (11) at (-2,1) {};
	    \node [style=Z dot] (12) at (-2,2) {};

	    \node [style=none] (20) at (-3,0) {};
	    \node [style=none] (21) at (-3,1) {};
	    \node [style=none] (22) at (-3,2) {};

	    \node [style=none] (30) at (5,0) {};
	    \node [style=none] (31) at (5,1) {};
	    \node [style=none] (32) at (5,2) {};

	    \node [style=none] (80) at (0,3) {{\scriptsize Frontier}};
	    \node [style=none] (81) at (-1.3,-1) {{\scriptsize Unextracted}};
	    \node [style=none] (82) at (3.75,-1) {{\scriptsize Extracted}};

	\end{pgfonlayer}
	\begin{pgfonlayer}{edgelayer}
	    \path[dotted pattern] (20) to (10);
	    \path[dotted pattern] (21) to (11);
	    \path[dotted pattern] (22) to (12);

	    \draw[style=hadamard edge] (10) to (0);
	    \draw[style=hadamard edge] (11) to (1);
	    \draw[style=hadamard edge] (12) to (2);

	    \draw[style=hadamard edge] (1) to (0);
	    \draw[style=hadamard edge] (1) to (10);

	    \draw (0) to (30);
	    \draw (1) to (31);
	    \draw (2) to (32);

	\end{pgfonlayer}
\end{tikzpicture}
    \caption{Spiders with a single input and output can be
    extracted directly.}
    \label{b}
  \end{subfigure}

  \begin{subfigure}[t]{0.48\textwidth}
    \centering
    \begin{tikzpicture}
	\begin{pgfonlayer}{nodelayer}
	    \node [style=Z dot] (0) at (0,0) {\tiny $\epsilon_{3}$};
	    \node [style=Z dot] (1) at (0,1) {\tiny $\epsilon_{2}$};
        \node [style=Z dot] (2) at (0,2) {\tiny $\epsilon_{1}$};
            \begin{scope}[on background layer]
                \node[draw, color=luxembourg red, fill=luxembourg red!10, thick, inner sep=2pt, fit=(0) (1) (2)] (3) {};
            \end{scope}
	    \node [style=hadamard] (4) at (3,2) {};
	    \node [style=Z phase dot] (5) at (2,2) {$\frac{\pi}{4}$};
	    \node [style=Z dot] (6) at (1,0) {};
	    \node [style=Z dot] (7) at (1,1) {};

	    \node [style=Z dot] (10) at (-2,0) {};
	    \node [style=Z dot] (11) at (-2,1) {};
	    \node [style=Z dot] (12) at (-2,2) {};

	    \node [style=none] (20) at (-3,0) {};
	    \node [style=none] (21) at (-3,1) {};
	    \node [style=none] (22) at (-3,2) {};

	    \node [style=none] (30) at (5,0) {};
	    \node [style=none] (31) at (5,1) {};
	    \node [style=none] (32) at (5,2) {};

	    \node [style=none] (80) at (0,3) {{\scriptsize Frontier}};
	    \node [style=none] (81) at (-1.3,-1) {{\scriptsize Unextracted}};
	    \node [style=none] (82) at (3.75,-1) {{\scriptsize Extracted}};

	\end{pgfonlayer}
	\begin{pgfonlayer}{edgelayer}
	    \path[dotted pattern] (20) to (10);
	    \path[dotted pattern] (21) to (11);
	    \path[dotted pattern] (22) to (12);

	    \draw[style=hadamard edge] (10) to (0);
	    \draw[style=hadamard edge] (11) to (1);
	    \draw[style=hadamard edge] (12) to (2);

	    \draw[style=hadamard edge] (1) to (10);

	    \draw[style=hadamard edge] (6) to (7);

	    \draw (0) to (30);
	    \draw (1) to (31);
	    \draw (2) to (32);

	\end{pgfonlayer}
\end{tikzpicture}
    \caption{Connections between frontier spiders are extracted by
    CZ gates.}
    \label{c}
  \end{subfigure}
  ~
  \begin{subfigure}[t]{0.48\textwidth}
    \centering
    \begin{tikzpicture}
	\begin{pgfonlayer}{nodelayer}
	    \node [style=Z dot] (0) at (0,0) {\tiny $\epsilon_{3}$};
	    \node [style=Z dot] (1) at (0,1) {\tiny $\epsilon_{2}$};
        \node [style=Z dot] (2) at (0,2) {\tiny $\epsilon_{1}$};
            \begin{scope}[on background layer]
                \node[draw, color=luxembourg red, fill=luxembourg red!10, thick, inner sep=2pt, fit=(0) (1) (2)] (3) {};
            \end{scope}
	    \node [style=hadamard] (4) at (4,2) {};
	    \node [style=Z phase dot] (5) at (3,2) {$\frac{\pi}{4}$};
	    \node [style=Z dot] (6) at (2,0) {};
	    \node [style=Z dot] (7) at (2,1) {};

	    \node [style=X dot] (8) at (1,0) {};
	    \node [style=Z dot] (9) at (1,1) {};

	    \node [style=Z dot] (10) at (-2,0) {};
	    \node [style=Z dot] (11) at (-2,1) {};
	    \node [style=Z dot] (12) at (-2,2) {};

	    \node [style=none] (20) at (-3,0) {};
	    \node [style=none] (21) at (-3,1) {};
	    \node [style=none] (22) at (-3,2) {};

	    \node [style=none] (30) at (5,0) {};
	    \node [style=none] (31) at (5,1) {};
	    \node [style=none] (32) at (5,2) {};

	    \node [style=none] (80) at (0,3) {{\scriptsize Frontier}};
	    \node [style=none] (81) at (-1.3,-1) {{\scriptsize Unextracted}};
	    \node [style=none] (82) at (3.75,-1) {{\scriptsize Extracted}};
	\end{pgfonlayer}
	\begin{pgfonlayer}{edgelayer}
	    \path[dotted pattern] (20) to (10);
	    \path[dotted pattern] (21) to (11);
	    \path[dotted pattern] (22) to (12);

	    \draw[style=hadamard edge] (10) to (0);
	    \draw[style=hadamard edge] (11) to (1);
	    \draw[style=hadamard edge] (12) to (2);

	    \draw[style=hadamard edge] (6) to (7);

	    \draw (8) to (9);

	    \draw (0) to (30);
	    \draw (1) to (31);
	    \draw (2) to (32);

	\end{pgfonlayer}
\end{tikzpicture}
    \caption{CNOT gates extract wires between frontier and
    unextracted spiders.}
    \label{d}
  \end{subfigure}
  \caption{The different steps of circuit extraction.}
  \label{fig:circ-extract}
\end{figure*}
Polynomial-time circuit extraction algorithm exists for ZX-diagrams that preserve causal flow, general flow or Pauli flow~\cite{duncanGraphtheoreticSimplificationQuantum2020,backensThereBackAgain2020,simmons2021Measurement}.
Duncan et al.~\cite{duncanGraphtheoreticSimplificationQuantum2020} introduce the first circuit extraction algorithm for ZX-diagrams with causal flow and focused general flow. 
Backens et al.~\cite{backensThereBackAgain2020} extend the previous work of Kissinger et al.~\cite{kissinger2019cnotcircuitextractiontopologicallyconstrained} to consider general flow.

This section introduces the principles of the circuit extraction algorithm for graph-like~(Definition~\ref{def:graph-like}) ZX-diagrams by Duncan et al.~\cite{duncanGraphtheoreticSimplificationQuantum2020}.
These techniques reemerge in various works other works, including the standard extraction algorithm by Backens et al.~\cite{backensThereBackAgain2020} and architecture-aware extraction (e.g., Villoria et al.~\cite{villoria2025optimizationsynthesisquantumcircuits}, Staudacher et al.\cite{staudacher2024neutralatom}, and Kissinger et al.~\cite{kissinger2019cnotcircuitextractiontopologicallyconstrained}).

Figure~\ref{fig:circ-extract} illustrates the main steps during circuit extraction.
The ZX-diagram is divided into an extracted and unextracted part separated by a frontier  $\epsilon_{1},\dots,\epsilon_{3}$.
The key to circuit extraction is that the addition of a CNOT gate changes the adjacency of the frontier.
Specifically, it adds or removes a Hadamard wire between a frontier and a neighboring spider depending on the placement of the control and target qubit.
This transformation does not change the underlying linear map as shown by Duncan et al.~\cite{duncanGraphtheoreticSimplificationQuantum2020}.

Starting from right to left (Figure~\ref{fig:adjacency-example}), circuit extraction is an iterative process until no unextracted part remains:
\begin{enumerate}
  \item Spiders that have a single input and output wire can be extracted directly by single qubit gates. 
   Figure~\ref{a} shows the extraction of a T-gate in the top row.
  \item The wire between two directly connected frontier spiders is extracted by a CZ gate $\left(\begin{ZX}
          \zxN{} \rar & [\zxwCol] \zxZ{} \dar[draw=blue, thick, densely dotted] \rar & [\zxwCol] \zxN{} \\
\zxN{} \rar & \zxZ{} \rar & [\zxwCol] \zxN{} \\
\end{ZX}\right)$.
        Figure~\ref{b} shows the CZ gate that
    results from the extraction of the directly connected frontier
    spiders $\epsilon_{2}$ and $\epsilon_{3}$.
\item CNOT $\left(\begin{ZX}
\zxN{} \rar & [\zxwCol] \zxZ{} \dar \rar & [\zxwCol] \zxN{} \\
\zxN{} \rar & \zxX{} \rar & [\zxwCol] \zxN{} \\
\end{ZX}\right)$ gates are used for frontier spiders that are connected
        to multiple unextracted spiders. Figure~\ref{c} shows how a CNOT
    gate is added to extract the diagonal connection of $\epsilon_{2}$.
\end{enumerate}

\begin{figure}[tb!]
    \centering
        \begin{tikzpicture}
	\begin{pgfonlayer}{nodelayer}
	    \node [style=Z dot] (0) at (0,0) {$\epsilon_{3}$};
	    \node [style=Z dot] (1) at (0,1) {$\epsilon_{2}$};
        \node [style=Z dot] (2) at (0,2) {$\epsilon_{1}$};
	    \node [style=hadamard] (4) at (3,2) {};
	    \node [style=Z phase dot] (5) at (2,2) {$\frac{\pi}{4}$};
	    \node [style=Z dot] (6) at (1,0) {};
	    \node [style=Z dot] (7) at (1,1) {};

        \node [style=Z dot] (10) at (-2,0) {$v_{3}$};
	    \node [style=Z dot] (11) at (-2,1) {$v_{2}$};
	    \node [style=Z dot] (12) at (-2,2) {$v_{1}$};

	    \node [style=none] (20) at (-3,0) {};
	    \node [style=none] (21) at (-3,1) {};
	    \node [style=none] (22) at (-3,2) {};

	    \node [style=none] (30) at (5,0) {};
	    \node [style=none] (31) at (5,1) {};
	    \node [style=none] (32) at (5,2) {};

	    \node [style=none] (80) at (0,3.75) {{Frontier}};
	    \node [style=none] (81) at (-1.3,-1.75) {{Unextracted}};
	    \node [style=none] (82) at (3,-1.75) {{Extracted}};

        \node (100) at ($(current bounding box.south)!0.5!(current bounding box.north) + (8,0)$) {$\mat{A}=
        \begin{bNiceMatrix}[first-row,first-col]
            & v_1 & v_2 & v_3 \\
          \epsilon_1 & 1 & 0 & 0 \\
          \epsilon_2 & 0 & 1 & 1 \\
          \epsilon_3 & 0 & 0 & 1 \\
        \end{bNiceMatrix}
        $};
	\end{pgfonlayer}
	\begin{pgfonlayer}{edgelayer}
	    \path[dotted pattern] (20) to (10);
	    \path[dotted pattern] (21) to (11);
	    \path[dotted pattern] (22) to (12);

	    \draw[style=hadamard edge] (10) to (0);
	    \draw[style=hadamard edge] (11) to (1);
	    \draw[style=hadamard edge] (12) to (2);

	    \draw[style=hadamard edge] (1) to (10);

	    \draw[style=hadamard edge] (6) to (7);

	    \draw (0) to (30);
	    \draw (1) to (31);
	    \draw (2) to (32);

	\end{pgfonlayer}

    \begin{pgfonlayer}{background}
        \node[draw, color=luxembourg blue, line width=2pt,style=dashed, inner sep=10pt, fit=(0) (1) (2) (10) (11) (12)] (91) {};
        \node[draw, color=luxembourg red, fill=luxembourg red!10, thick, inner sep=2pt, fit=(0) (1) (2)] (3) {};
        % --- fade everything OUTSIDE node 91 ---
        \fill[black, opacity=.15, even odd rule]
        ([xshift=-2pt, yshift=-2pt]current bounding box.south west) rectangle ([xshift=2pt, yshift=2pt]30.east |- current bounding box.north east)
(91.south west) rectangle (91.north east);
    \end{pgfonlayer}
\end{tikzpicture}
    \caption{Example adjacency matrix between the frontier spiders $\epsilon_{1},\dots,\epsilon_{3}$ and nearest unextracted spiders $v_{1},\dots,v_{3}$.}
    \label{fig:adjacency-example}
\end{figure}

Frontier spiders with multiple connected neighbors are extracted based on their binary adjacency matrix.
A spider is directly extractable iff it connects to a single frontier spider, indicated by rows contain a single non-zero entry (rule 1).
The frontier can be manipulated by row additions of its adjacency matrix, with the goal to eliminate edges until rule 1 is applicable.
Duncan et al.~\cite{duncanGraphtheoreticSimplificationQuantum2020} showed that the addition of two rows corresponds to their bitwise XOR, with each row addition adding a CNOT gate and removing or adding a Hadamard wire.
%Each row addition introduces one CNOT gate in the extracted circuit.

Let's consider the example shown in Figure~\ref{fig:adjacency-example} that admits the adjacency matrix $\mat{A}$ for frontier spiders $\epsilon_{1},\dots,\epsilon_{3}$ and neighbors $v_{1},\dots,v_{3}$ (Figure~\ref{c}), where $\mat{A}_{\epsilon_{i}v_{j}} = 1$ iff $\epsilon_i$ connects $v_{j}$, and $0$ otherwise.
In this example, the addition of the last two rows adds a Gaussian elimination gate with control $\epsilon_{2}$ and target $\epsilon_{3}$ qubit (Figure~\ref{c}). 

The state-of-the-art extraction algorithm uses Gaussian elimination to solve the adjacency matrix restricted to row additions and row swaps~\cite{duncanGraphtheoreticSimplificationQuantum2020,backensThereBackAgain2020}.
Anton~\cite{anton2013elementaryalgebra} offers a detailed introduction to the Gaussian elimination algorithm.
A key limitation is that the Gaussian elimination algorithm does not solve the adjacency matrix with the minimal number of row additions~\cite{strassen1969gaussianelimination,macedo2016gaussianeliminationrevisited}.
Consequently, reducing the number of row additions decreases the overall CNOT gate count of the extracted circuit.

\section{Target Metrics}
\label{sec:target_metrics}
%%\begin{itemize}
%%\item \xxx{identify the best method per metric and mention that}
%%\item \xxx {Add table that contains the different algorithms, circuit types and metrics; Benchmark}
%%\end{itemize}
%
ZX-calculus allows for architecture-independent and architecture-aware optimization strategies. 
The objective of quantum circuit optimization is to reduce the resource demand of quantum programs and enable their execution on real quantum devices.
\subsection{Noisy Intermediate-Scale Quantum Computing}
Current NISQ-era devices are resource restricted and error-prone, primarily due to noise forcing decoherence~\cite{preskill2018nisq}.
They are limited in total gate count, circuit depth, two-qubit gate count, and qubit connectivity.
In order to produce an executable quantum circuit, architectural characteristics need to be incorporated.
For many NISQ-era quantum architectures, two-qubit gates contribute significantly more noise than typical single-qubit Clifford gates~\cite{preskill2018nisq}.
These noisy gates often require sophisticated error correction that further increases the demand for resources (e.g., Shor~\cite{shor1995qec}, Steane~\cite{steane1996qec}, and topological codes~\cite{bravyi1998qec}).

\paragraph{Circuit Depth}
Circuit depth is directly linked to the execution time and noise exposure of the circuit.
Shallower circuits can be prepared and executed faster. 
A quantum circuit can only be executed reliably on a quantum device if its execution time is shorter than the coherence time (see Section~\ref{sec:fundamentals}).
The coherence time and gate error rates vary across architectures.

\paragraph{Two-Qubit Gate Count}
ZX-diagrams consist of generators and not quantum gates. 
The extraction of a quantum circuit is not trivial and might negate optimization success at the ZX-diagram level (see Section~\ref{sec:circuit-extraction-problem}).
Some works approximate the number of two-qubit gates from the number of Hadamard wires contained in ZX-diagrams as proposed by~\citet{staudacherReducing2QuBitGate2023, holkerCausalFlowPreserving2024}.
Architecture-aware strategies need to take hardware-dependent factors into account, such as qubit connectivity~\cite{kissinger2019cnotcircuitextractiontopologicallyconstrained}.

\subsection{Fault-Tolerant Quantum Computing}
Fault-tolerant quantum computing is the key to large-scale quantum computing.
It implements logical qubits with many physical qubits and error-correction, such that errors can be corrected faster than they accumulate~\cite{shor1996faulttolerance,gottesman2009qec}.

\paragraph{Qubit Count}
Quantum circuits are composed of quantum gates that act on logical qubits.
In the context of ZX-diagrams, the number of logical qubits is the maximum number of wires intersected by any vertical cut through the diagram~\cite{vandaele2024qubitcountoptimizationusingzxcalculus}.

\paragraph{T-Gate Count}
The T-gate is a particularly noisy single-qubit gate that forms a universal gate set when combined with the Clifford gate set. 
Unlike typical Clifford gates, the T-gate significantly adds quantum error correction overhead on certain fault-tolerant quantum computing architectures~\cite{fowlerSurfaceCodesPractical2012}. 
Consequently, a prime objective of architecture-independent optimization is to reduce the T-gate count.

\section{Survey}
\label{sec:taxonomy}
This survey categorizes ZX-based quantum circuit optimization algorithms by their underlying strategy and target metric.
We selected 26 works with optimization strategies that focus on the semantic-preserving rewriting rules or improve the circuit extraction step.

Each approach is classified by its main optimization algorithm
If an entry contains "Ad-hoc", the respective work introduces a novel procedure that reduces the indicated metric. 
If a heuristic is indicated, the optimization is guided by characteristics of the ZX-diagram.
Other entries specify the algorithms used such as simulated annealing (SA), reinforcement learning (RL), genetic algorithms (GA), look-ahead (LA), integer linear programming (ILP), directed vertex feedback solver (DVFS) and template matching (TM).

Figure~\ref{fig:survey-overview} shows the focus of the surveyed works.
It counts the number of works that target a specific metric for each optimization strategy.
An optimizer contributes a count of one to every target metric in the figure.
In case a method approximates quantum circuit metrics at the ZX-diagram level, both are indicated.
For example, this is the case when the number of two-qubit gates is approximated by the number of Hadamard wires.

Table~\ref{tabsurvey} provides a high-level overview of our survey.
We differentiate between architecture-agnostic in blue and architecture-aware strategies in red.

\begin{figure}[tb!]
    \begin{center}
        \begin{tikzpicture}
          \begin{axis}[
              width=0.8\columnwidth, height=6cm,
              xlabel={Metric}, ylabel={Algorithm Category},
              xtick={1,2,3,4,5,6},
              ytick={1,2,3,4,5},
              xticklabels={$T$,$2Q$,$d$,$e$,$v$,$Q$},
              yticklabels={Ad-hoc, Heuristic, Search, Stochastic, RL},
              x tick label style={rotate=0, anchor=east},
              grid=both,
              ylabel style={at={(axis description cs:-0.12,0.5)}, anchor=south},
              xticklabel style={yshift=-12pt, anchor=base},
            ]
            % coordinates: (metricIndex, algIndex) [size]
            \addplot[
              only marks,
              scatter,
              scatter src=explicit,
              mark=*,
              visualization depends on={\thisrow{size} \as \s},
              scatter/@pre marker code/.append style={
                /tikz/mark size={2.5 + 1*\s}
              },
              scatter/@post marker code/.append code={
                \node[font=\scriptsize, text=white] at (0,0)
                {\pgfmathprintnumber[precision=0]{\pgfplotspointmeta}};
              },
            ] table[meta=size, row sep=\\] {
              x y size\\
              %% Ad-hoc
              1 1 8\\
              2 1 7\\
              3 1 4\\
              5 1 1\\
              6 1 1\\
              %% Heuristics
              1 2 4\\
              2 2 5\\
              3 2 1\\
              4 2 2\\
              6 2 1\\
              %% Search
              1 3 1\\
              2 3 3\\
              3 3 1\\
              4 3 2\\
              6 3 1\\
              %% Stochastic
              2 4 5\\
              3 4 3\\
              4 4 2\\
              %% RL
              1 5 1\\
              2 5 1\\
              4 5 2\\
              5 5 1\\
            };
          \end{axis}
      \end{tikzpicture}
    \end{center}
    \caption{Number of works classified by their optimization strategy and target metric.}
    \label{fig:survey-overview}
\end{figure}

The following sections discuss the individual approaches and aim to highlight connections and identify future research directions.

\begin{table*}[p]
  %\begin{sidewaystable}
  \setlength\tabcolsep{0pt}
  {\footnotesize
    \begin{tabular*}{\textwidth}%{lclcccccc}
      {@{\extracolsep{\fill}}
        >{\columncolor{white}[0pt][5pt]}l
        >{\columncolor{white}[5pt][5pt]}c
        >{\columncolor{white}[5pt][5pt]}l
        >{\columncolor{white}[5pt][5pt]}c
        >{\columncolor{white}[5pt][5pt]}c
        >{\columncolor{white}[5pt][5pt]}c
        >{\columncolor{white}[5pt][5pt]}c
        >{\columncolor{white}[5pt][5pt]}c
        >{\columncolor{white}[5pt][0pt]}c
      @{}}
      \toprule
      \textit{Author(s)} & \textit{Year} & \textit{Algorithm} &
      \textit{T} & \textit{2Q} & \textit{Depth} & \textit{Qubits} &
      \textit{Edges}  & \textit{Vertices} \\
      \midrule
      \rowcolor{luxembourg blue!20} \citet{faganOptimisingCliffordCircuits2019} & 2019 &
      Ad-hoc & &  \yestaxonomy & \yestaxonomy & & & \\
      \rowcolor{luxembourg red!20} \citet{kissinger2019cnotcircuitextractiontopologicallyconstrained}
      & 2019 & Ad-hoc & & \yestaxonomy & & & & \\
      \rowcolor{luxembourg blue!20} \citet{beaudrap2020org} & 2020 & Ad-hoc + Heuristic &
      \yestaxonomy& & & & &  \\
      \rowcolor{luxembourg blue!20} \citet{beaudrap2020extended} & 2020 & Ad-hoc + Heuristic &
      \yestaxonomy& & & & & \\
      \rowcolor{luxembourg blue!20} \citet{duncanGraphtheoreticSimplificationQuantum2020} & 2020 &
      Ad-hoc & \yestaxonomy & & & & & \\
      \rowcolor{luxembourg blue!20} \citet{kissinger2020journal} & 2020 &
      Ad-hoc & \yestaxonomy & & & & & \\
      \rowcolor{luxembourg red!20} \citet{cowtanPhaseGadgetSynthesis2020} & 2020 & Ad-hoc +
      Greedy & & \yestaxonomy & & & & \\
      \rowcolor{luxembourg red!20} \citet{cowtan2020genericcompilationstrategyunitary} & 2020 & Ad-hoc &
      & \yestaxonomy &  & & & \\
      \rowcolor{luxembourg blue!20} \citet{munson2021And}
      & 2021 & Ad-hoc + Heuristic & \yestaxonomy& & & & & \\
      %\rowcolor{luxembourg red!20} Borgna et
      %al.~\cite{borgnaHybridQuantumclassicalCircuit2021a} & 2021 &
      %Ad-hoc & \yestaxonomy & & & & & \\
      \rowcolor{luxembourg red!20} \citet{zilk2022photonic} & 2022 &
      Ad-hoc & \yestaxonomy & & & & & \yestaxonomy\\
      \rowcolor{luxembourg red!20} \citet{gogiosoAnnealingOptimisationMixed2023} & 2023 & SA &
      & \yestaxonomy & & & & \\
      \rowcolor{luxembourg blue!20} \citet{staudacherReducing2QuBitGate2023} & 2023 & Rand. +
      Heuristic & &  \yestaxonomy & & & \yestaxonomy & \\
      \rowcolor{luxembourg red!20} \citet{Winderl_2023} &
      2023 & Ad-hoc + Heuristic & & \yestaxonomy & & & &  \\
      \rowcolor{luxembourg blue!20} \citet{riuReinforcementLearningBased2023} & 2023 & RL & &
      \yestaxonomy & & & \yestaxonomy & \\
      \rowcolor{luxembourg red!20} \citet{Meijer_van_de_Griend_2023} & 2023 & Ad-hoc +
      Heuristic & & & \yestaxonomy & &  & \\
      \rowcolor{luxembourg blue!20} \citet{vandaele2024qubitcountoptimizationusingzxcalculus}
      & 2024 & Ad-hoc + DVFS + Greedy & & & & \yestaxonomy & & \\
      \rowcolor{luxembourg blue!20} \citet{holkerCausalFlowPreserving2024} & 2024 & Rand. +
      Heuristic & &  \yestaxonomy & & &  \yestaxonomy & \\
      \rowcolor{luxembourg blue!20} \citet{nageleOptimizingZXDiagramsDeep2023} & 2024 & RL &
      & & & & & \yestaxonomy \\
      \rowcolor{luxembourg red!20} \citet{staudacher2024neutralatom} & 2024 & Ad-hoc & &  &
      \yestaxonomy & & & \\
      \rowcolor{luxembourg red!20} \citet{ewen2025genetic} &
      2025 & GA & &  \yestaxonomy & \yestaxonomy & &  & \\
      \rowcolor{luxembourg red!20} \citet{huang2024redefininglexicographicalorderingoptimizing} &
      2025 & Ad-hoc + Heuristic & \yestaxonomy & \yestaxonomy & & & & \\
      %\rowcolor{luxembourg red!20}
      % Heavy~\cite{heavey2025improvedtcountsactive} & 2025 & &
      % \yestaxonomy & & & & & \\
      \rowcolor{luxembourg blue!20} \citet{mattick2025optimizingquantumcircuitszx} & 2025 & RL +
      Tree search & &  \yestaxonomy & & & \yestaxonomy & \\
      \rowcolor{luxembourg blue!20} \citet{fischbach2025exhaustivesearchquantumcircuit} & 2025 &
      Tree search & \yestaxonomy & & & & \yestaxonomy & \\
      \rowcolor{luxembourg red!20} \citet{liu2024} & 2024 &
      TM & & \yestaxonomy & \yestaxonomy & & & \\
      %\rowcolor{luxembourg red!20} Liu et al.~\cite{liu2025} & 2025 &
      %\gls{tm} + \gls{ga} & & \yestaxonomy & \yestaxonomy & & & \\
      %\rowcolor{luxembourg blue!20} Kelly \& Kissinger~\cite{kelly2025}
      %& 2025 & & & & & & & \\
      \rowcolor{luxembourg blue!20} \citet{chen2025quantumcircuitoptimizationbased} & 2025 &
      SA + LA & &  \yestaxonomy & \yestaxonomy & & & \\
      \rowcolor{luxembourg red!20} \citet{villoria2025optimizationsynthesisquantumcircuits} & 2025 &
      ILP + Peephole & &  \yestaxonomy & & & & \\
      \bottomrule
  \end{tabular*}}
  \caption{Overview of the different quantum circuit optimization
    algorithms, the optimization procedure and the main target metrics.
    Architecture-independent approaches are highlighted in blue and
  architecture-aware approaches are highlighted in red.}
  \label{tabsurvey}
  %\end{sidewaystable}
\end{table*}
\section{Target Metrics}
\label{sec:target_metrics}
%\begin{itemize}
%\item \xxx{identify the best method per metric and mention that}
%\item \xxx {Add table that contains the different algorithms, circuit types and metrics; Benchmark}
%\end{itemize}

ZX-calculus allows for architecture-independent and architecture-aware optimization strategies. 
The objective of quantum circuit optimization is to reduce the resource demand of quantum programs and enable their execution on real quantum devices.

\subsection{Noisy Intermediate-Scale Quantum Computing}
Current NISQ-era devices are resource restricted and error-prone, primarily due to noise forcing decoherence~\cite{preskill2018nisq}.
They are limited in total gate count, circuit depth, two-qubit gate count, and qubit connectivity.
In order to produce an executable quantum circuit, architectural characteristics need to be incorporated.
For many NISQ-era quantum architectures, two-qubit gates contribute significantly more noise than typical single-qubit Clifford gates~\cite{preskill2018nisq}.
These noisy gates often require sophisticated error correction that further increases the demand for resources (e.g., Shor~\cite{shor1995qec}, Steane~\cite{steane1996qec}, and topological codes~\cite{bravyi1998qec}).

\paragraph{Circuit Depth}
Circuit depth is directly linked to the execution time and noise exposure of the circuit.
Shallower circuits can be prepared and executed faster. 
A quantum circuit can only be executed reliably on a quantum device if its execution time is shorter than the coherence time (see Section~\ref{sec:fundamentals}).
The coherence time and gate error rates vary across architectures.

\paragraph{Two-Qubit Gate Count}
ZX-diagrams consist of generators and not quantum gates. 
The extraction of a quantum circuit is not trivial and might negate optimization success at the ZX-diagram level (see Section~\ref{sec:circuit-extraction-problem}).
Some works approximate the number of two-qubit gates from the number of Hadamard wires contained in ZX-diagrams as proposed by~\citet{staudacherReducing2QuBitGate2023, holkerCausalFlowPreserving2024}.
Architecture-aware strategies need to take hardware-dependent factors into account, such as qubit connectivity~\cite{kissinger2019cnotcircuitextractiontopologicallyconstrained}.

\subsection{Fault-Tolerant Quantum Computing}
fault-tolerant quantum computing is the key to large-scale quantum computing.
It implements logical qubits with many physical qubits and error-correction, such that errors can be corrected faster than they accumulate~\cite{shor1996faulttolerance,gottesman2009qec}.

\paragraph{Qubit Count}
 quantum circuits are composed of quantum gates that act on logical qubits.
In the context of ZX-diagrams, the number of logical qubits is the maximum number of wires intersected by any vertical cut through the diagram~\cite{vandaele2024qubitcountoptimizationusingzxcalculus}.

\paragraph{T-Gate Count}
The T-gate is a particularly noisy single-qubit gate that forms a universal gate set when combined with the Clifford gate set. 
Unlike typical Clifford gates, the T-gate significantly adds quantum error correction overhead on certain fault-tolerant quantum computing architectures~\cite{fowlerSurfaceCodesPractical2012}. 
Consequently, a prime objective of architecture-independent optimization is to reduce the T-gate count.

\section{Heuristic}

\subsection{Circuit Depth}
\citet{faganOptimisingCliffordCircuits2019}~introduced the first ZX-based quantum circuit optimization algorithm.
Their optimization strategy moves Pauli gadgets towards the inputs and group CNOT and TONC gates (CNOT gates with switched control and target qubits) that act on the same qubits.
They demonstrate a significant reduction of the CNOT gate count by $\approx 16\%$ and circuit depth up to $\approx 30 \%$ for pure Clifford circuits.
%\todo{Add missing publications}

\paragraph{Architecture-Aware}

\citet{staudacher2024neutralatom}~add a step to the circuit extraction algorithm~\citet{backensThereBackAgain2020} between the CNOT and single-qubit gate extraction to target neutral-atom architectures.
Instead of using the Clifford + T gate set, this work targets a gate
set that consists of Z gates, multi-controlled phase gates and global single qubit rotations in the XY-plane.
Phase gadgets are the native representation of multi-controlled phase gates.
For identified frontier phase gadgets, the fusion rule and reverse
gadget fusion rules are applied. Potentially missing phase gadgets
are added to extract a full multi-controlled phase gate.
Neutral-atom quantum computers can natively execute multi-controlled
phase gates and are limited by execution time and the total gate count.
Especially the global single-qubit rotation is an order of magnitude
slower than the multi-controlled phase gate and the Z phase gates.
The circuit's execution time is computed along the circuit as the sum
of individual gates execution time, which is assumed to increase with
the rotation angle.
Their approach outperforms the execution time of Qiskit compiled
circuits from $26\%$ up to $40\%$, primarily due to the reduction in
the number of global phase gates.

\subsection{T-Gate Count}
\paragraph{Architecture-Independent}
\citet{kissinger2020journal}~use the theoretical framework~\citet{duncanGraphtheoreticSimplificationQuantum2020} to allow
semantic preserving optimization of circuits composed of the entire
Clifford + T-gate set.
Their procedure aims to minimize the T-gate count.
Their simplification strategies apply local complementation and
pivoting to remove Pauli spiders, spiders with a phase that is a
multiple of $\pi$, at the expense of adding phase
gadgets. Applying the identity-removal and gadget fusion rules
efficiently remove the phase gadgets.
By rerunning the previous algorithm symbolically, it is possible to
identify non-local phase recombination that further reduce the T-gate count.
This step is known as \textit{phase teleportation}.
The resulting optimization algorithm is implemented as the standard
simplification routine in the PyZX library~\cite{kissingerPyZXLargeScale2020} under the name "full
reduce". It is found in many optimization pipelines as a pre- or
post-processing step for T-gate minimization.
The evaluation of a benchmark set that consists of 36 structured
standard quantum
circuits shows that "full reduce" improves the state-of-the-art
T-count on 17\% and matches on 72\% of the instances.
Further improvements were achieved when full reduce is paired with
the dedicated T-gate optimizer TODD~\cite{Heyfron2019}. Combining
both methods improve the T-count in 56 \% of the
benchmark circuits.

The following approaches develop spider nest identities for
$\frac{\pi}{4}$-parity-phase operations to eliminate T-gates from  quantum circuits~\cite{beaudrap2020org}. A spider nest is a ZX-diagram that consists only of $\frac{\pi}{4}$ phase gadgets.
Instead of rules that modify a single or few generators, spider nest
identities take the full nest into account.
\citet{munson2021And}~independently derived the same spider nest identities
and generalized them by connecting ZX-calculus and logical AND
gates.

\citet{beaudrap2020org, beaudrap2020extended}~introduce the phase gadget elimination (PHAGE)
strategy, a systematic method
that takes advantage of spider nest identities to reduce the T-gate
count in ZX-diagrams.
The core idea of PHAGE is to decompose larger and more complex phase
gadgets into simpler and smaller ones. This decomposition allows for
the effective application
of spider nest identities to eliminate T-gates.
The evaluation of PHAGE on a benchmark set of 35 parameterized
circuits reveals that the state-of-the-art T-gate counts are improved
for 21 instances under runtime constraints.

\paragraph{Architecture-Aware}
\citet{zilk2022photonic}~use the phase teleportation algorithm~\citet{kissinger2020journal} to reduce the T-gate count and the Clifford elimination algorithm of~\citet{faganOptimisingCliffordCircuits2019} to target photonic quantum computers.
The first step in their procedure is to express a quantum circuit only as a sequence of Hadamard, CZ, and arbitrary Z-phase gates.
The resulting circuit is converted to a measurement graph and interpreted as a graph-like~(Definition~\ref{def:graph-like}) ZX-diagram.
This ZX-diagram is optimized with the phase teleportation and Clifford elimination algorithms.
In the MBQC framework, every spider directly corresponds to a qubit; hence a reduction of spiders (e.g. Clifford and T-spiders) decreases the number of qubits.
Measurement-graphs can be directly converted into hardware instructions. 
The limiting factor of photonic architectures is the number of available photons and optical instruments.
Compared to the standard photonic compiler Perceval~\cite{heurtel2023perceval}, the ZX-based approach successfully compiles all benchmarked circuits and mostly outperforms it with respect to the photon count and optical instrument count.

\subsection{Qubit Count}
\citet{vandaele2024qubitcountoptimizationusingzxcalculus}~introduced an approach to minimize the logical qubit
count in ZX-diagrams by systematically reordering and (un)bending
spiders. The
core idea is to use the rewrite rules to change the structure of the
ZX-diagram such that it is possible to find a vertical cut that
reduces the maximum number of wires, hence minimizing the logical qubit count.
The NP-hard problem of minimizing the number of wires through vertical cuts is formerly known as the "fixed-endbags path width problem".
Consequently, finding an optimal solution is computationally
expensive for large diagrams.

\subsection{Two-Qubit Gate Count}
\paragraph{Architecture-Independent}
\citet{staudacherReducing2QuBitGate2023}~demonstrate that an average reduction of the edge
count by 23\% translates to a two-qubit gate
reduction of 16\% by applying a set of general flow-preserving rewrite
rules (identity removal, spider fusion, local complementation, and
pivoting).
They introduce a new heuristic based
on the expected change of the Hadamard wire count by the local
complementation and pivoting
rules.
A reduction of the Hadamard wire count correlates with the expected
reduction of two-qubit gates. Slightly worsening of the cost function
was allowed to permit further improvements later on.
Stochastic and greedy algorithms are used to select the rewriting rules.
Further integration with the NAM framework allowed for greater edge
count reductions of 29\% that translated
into two-qubit gate reductions of approximately 21\%.

In a related work, \citet{holkerCausalFlowPreserving2024}~extends these ideas by focusing on
diagrams that preserve the stricter causal flow
property.
If causal flow is preserved, the effect of a rewriting rule on the two-qubit
gate count can be exactly quantified from change in the wire count,
completely bypassing circuit extraction for all intermediate
optimization steps.
By maintaining causal flow, \citet{holkerCausalFlowPreserving2024}~achieves an average two-qubit gate count
reduction of 20\%.
ZX-diagrams with causal flow have a circuit-like structure, that allows for a
straightforward and efficient extraction.
Moreover, verifying causal flow is computationally less expensive than verifying
general flow. However, it is important to note that only a limited set of
rewriting rules, namely identity removal and spider fusion, are
known to preserve causal flow. This limitation severely restricts the
possible diagram transformations but offers a trade-off between
the solution quality and verification complexity.

The following two optimization strategies use the
tket~\cite{Sivarajah2021} compiler.
\citet{cowtanPhaseGadgetSynthesis2020}~demonstrated that the efficient pairwise synthesis of
Pauli gadgets using tket decreased the average CNOT gate count by
$\approx 55 \%$ and improved the two-qubit depth
by $\approx 58 \%$.
In a subsequent work, \citet{cowtan2020genericcompilationstrategyunitary}~showed that a three-step
optimization routine improves the CNOT gate count and depth of the unitary coupled cluster (UCC) Ansatz, a subroutine of variation quantum eigensolver (VQE), by $69\%$ and $75\%$:
(i) the partition of the ZX-diagram of the UCC Ansatz into commuting
sets is treated as a graph coloring problem, (ii) the resulting
Pauli gadgets are converted to Phase gadgets, and (iii) the phase
gadgets are efficiently synthesized using Matroid partitioning~\cite{amyPolynomialTimeTDepthOptimization2014}.

Another promising approach to quantum circuit optimization is the
aggregation of multiple subcircuits that can be reordered and
replaced by optimized template circuits. Template circuits are
pre-optimized subcircuits that implement the same program flow as the
subcircuit they are meant to substitute. \citet{liu2024}~introduce a
string-based intermediate representation of subcircuits and a
template matching algorithm that improves the CNOT gate
count. Although not explicitly relying on rewriting
rules for optimization, \citet{liu2024} employ ZX-calculus to verify the
correctness of the intermediate representation and their associated
templates.

%Building upon this framework, it would be possible to
%create optimal templates for small subcircuits, potentially in
%parallel, using ZX-calculus and iteratively building larger templates
%and circuits.

\paragraph{Architecture-Aware}
\citet{kissinger2019cnotcircuitextractiontopologicallyconstrained}~introduce the first architecture-aware optimization algorithm and
outperform the existing compiler frameworks
QuilC~\cite{smith2017practicalquantuminstructionset} and
tket~\cite{cowtan2019qubitrouting}.
During circuit extraction, they restrict the Gaussian elimination algorithm of
the parity map to only include nearest-neighbor rows using the
Steiner-Gauss algorithm.
Consequently, CNOT gates can only act between neighboring qubits.

\citet{villoria2025optimizationsynthesisquantumcircuits}~modified the circuit extraction algorithm
of~\citet{backensThereBackAgain2020} to target trap-ion
computers that are highly dependent on global
gates.
Instead of Gaussian elimination algorithm, the frontier is extracted solving a
linear program that ensures only vertices are extracted which can
be included in the same global gate.
Afterward peephole optimization merges the extracted two-qubit gates
into global GMS gates.
This algorithm outperforms the Qiskit implementation on most quantum circuits.

\citet{Meijer_van_de_Griend_2023}~develop a two-step recursive optimization strategy
that considers qubit connectivity restrictions of the underlying
quantum architecture, thus circumventing the need for an additional
routing step, by using the notion of phase
polynomials.
The algorithm introduces a biadjacency matrix $\mat P$ between the
phase gadgets
legs and the attached qubits.
For the base recursion step, the phase gadget
with the lowest connectivity that is not needed to synthesize other
phase gadgets is removed.
The selected phase gadget must be a non-cutting vertex, meaning the
column in $\mat P$ with the most $0$ or $1$ entries.
Afterward, the phase gadgets are synthesized by decomposition into a
CNOT ladder as
shown in Figure~\ref{figphasegadget}.
The second recursion step removes rows from the remaining
phase-gadget biadjacency matrix by row addition.
Every row operation adds CNOT gates to the ladder.
To eliminate excess CNOT gates, the Steiner-Gauss algorithm is run on
the resulting parity map.

%Instead of respecting
%the original circuit structure (e.g. using SWAP gates to follow the
%connectivity constraints), architecture-aware synthesis only
%preserves the underlying unitary operation.

\citet{winderlRecursivelyPartitionedApproach2023}~adapt the architecture-aware approach of~\citet{gogiosoAnnealingOptimisationMixed2023} by replacing
simulated annealing with a heuristic search in combination with a
divide-and-conquer
strategy.
The first step in their strategy is the simplification of the ZX
polynomial by removing, merging, and moving phase gadgets.
Afterward, the ZX-diagram is split into a left parity, a right
parity, and a ZX polynomial region.
A Gaussian elimination optimization algorithm minimizes the CNOT
count based on the combined cost of the region. The cost of removing
a phase gadgets
legs is computed by the minimal spanning tree of the architectural-dependent
connectivity.
Their novel heuristic based on the shortest path between the control
and target qubit of CNOT gates is used in conjunction with the
Steiner-Gauss algorithm to synthesize the phase gadgets in both parity regions.
The remaining ZX polynomial is regrouped from a leg-based score and,
following a divide-and-conquer strategy, split into subregions again.
These steps are recursively repeated until no ZX
polynomial remains.
Both
methods are outperformed by other state-of-the-art algorithms, such
as tket~\cite{Sivarajah2021}, for structured circuits.
However, the
heuristic approach
of~\citet{winderlRecursivelyPartitionedApproach2023}
exhibits better scaling in the qubit count and CNOT tree depth
compared to the stochastic approach by~\citet{gogiosoAnnealingOptimisationMixed2023}.

\citet{huang2024redefininglexicographicalorderingoptimizing}~develop a novel approach for architecture-aware
synthesis of Trotter
operators.
Trotterized time evolution operators can be expressed in terms of
Pauli gadgets. Pauli gadgets can be described by exponentiation of
Pauli strings.
Each letter of the Pauli string corresponds to a Pauli gate.
The core idea of their approach is to lexicographically reorder the
Pauli strings that compose the Pauli gadgets of the Trotter operator.
As a result, phase gadget legs with the same letter are grouped on
the same qubit.
Reordering of non-commuting Pauli gadgets is possible because the
introduced error (Trotter error) is outweighed by the error that
originates from noisy gates.
Their algorithm iteratively diagonalizes and disconnects qubits. The
highest entangled qubit is chosen for the current iteration.
The diagonalization step places single qubit Clifford gates on the
selected qubit until the gadget has either no leg or a Z leg.
Based on the Pauli gadget leg, the disconnection step introduces two
CNOT gates and up to two single-qubit Clifford gates.
This disconnection step reduces the entanglement of the selected qubit.
The selection of the qubit is determined by an entanglement heuristic.
Qubit entanglement is calculated from the occurrences of the I-gate in
the Pauli string and the length difference between the largest and
smallest substrings that exclude the I-gate.
This approach outperforms the state-of-the-art CNOT count for random
circuits and larger couple-cluster unitaries.

\section{Metaheuristic}
\subsection{Two-Qubit Gate Count}
\paragraph{Architecture-Independent}
\citet{chen2025quantumcircuitoptimizationbased}~improve the two-qubit gate count by $2\%$ compared to the
heuristic approach of~\citet{staudacherReducing2QuBitGate2023} using simulated annealing
to partition a quantum circuit into subcircuits that are
optimized.
In a first step, the quantum circuit is divided into sequential
layers of gates. Per layer, one single-qubit gate, one target and control qubit
of multi-qubit gates can act at most on each qubit.
Subcircuits are groups of sequential layers.
Starting from a random configuration of subcircuits, each circuit is
converted into a ZX-diagram, optimized, and extracted.
The resulting circuits are iteratively merged.
Following a delayed gate approach, known gate commutation and
substitution rules are used to further improve the circuit depth
and gate count. The delayed gate approach is implemented in
PyZX~\cite{kissingerPyZXLargeScale2020} under the name \textit{basic
optimization}.
For each iteration of the simulated annealing algorithm, the starting
configuration of the subcircuits is changed.

\paragraph{Architecture-Aware}
\citet{gogiosoAnnealingOptimisationMixed2023}~reduce the CNOT count for mixed
phase gadgets ZX-diagrams by $27\%$ on
a grid-shaped topology.
Their technique is grounded in the decomposition of phase gadgets
and is paired with simulated annealing and CNOT conjugation rules.
The ZX-diagram is split up into three layers: (i) a left
nearest-neighbor CNOT layer, (ii) a right nearest-neighbor CNOT
layer, and (iii) a mixed phase gadget ZX-diagram.
The cost of implementing a phase gadget is computed from the distance
between distinct legs that are mapped onto the topology.
This corresponds to the nearest-neighbor CNOT count.
The total cost of a circuit is the sum of all its CNOT gates.
Simulated annealing is used to explore different phase gadget mappings.
To exploit symmetries, the phase gadgets are converted to CNOT ladders
(Figure~\ref{figphasegadget}), so that gate conjugation rules
can be applied.

\citet{ewen2025genetic}~introduce a genetic programming approach for
synthesizing shallow quantum circuits with fewer two-qubit gates from
ZX-diagrams. The original quantum circuit is
converted into a ZX-diagram that will be evolved via a set of
mutation operations.
In their work, they implement two different categories of mutations.
Semantics-preserving mutations are formed by the rewriting rules of ZX-calculus.
Semantics-breaking mutations, such as edge flipping and edge addition,
introduce new connectivity patterns in the ZX-diagram. Although these
semantics-breaking mutations violate the correctness of the circuit, they
permit the exploration of a larger state-space, at the cost of
circuit fidelity.
Experimental results demonstrate that ZX-based genetic programming
can produce well-balanced circuit solutions, achieving results close
to the state-of-the-art for circuit depth, circuit fidelity, and
two-qubit gate count.

%Future research directions could include more
%architecture-specific constraints, such as hardware connectivity or
%gate error rate.

\section{Reinforcement Learning}
\subsection{Metric Agnostic}
\citet{nageleOptimizingZXDiagramsDeep2023}~implement a scalable and general reinforcement learning framework
for the optimization of ZX-diagrams that can be adapted for different
metrics.
Actions are grouped by node and edge impact, irrespective of the
preservation of general flow or causal flow.
As a consequence, quantum circuit metrics are not considered because
of the lack of circuit extraction.
In contrast to the reinforcement learning-approaches of~\citet{riuReinforcementLearningBased2023}
and~\citet{mattick2025optimizingquantumcircuitszx}, the
reversibility of rewriting rules is considered beyond spider
un/fusion by including the bialgebra rule.
Furthermore, the agent can mask actions to improve the efficiency of training.
%This framework can be extended to include more reversible rules,
%metrics, flow verification, and circuit extraction.

\subsection{Two-Qubit Gate Count}
\citet{riuReinforcementLearningBased2023}~introduce a reinforcement learning-based approach
that uses graph neural networks to minimize the two-qubit gate
count. They
restrict the set of rewriting rules to general flow-preserving
transformations. Circuit extraction is treated as a
black-box, simplifying the optimization pipeline.
The reinforcement learning agent is trained to either select and apply a rewriting rule
or to terminate the optimization process. The agent can apply
multiple rules until it decides to terminate. A reward function guides
the decision based on metrics such as the two-qubit gate count.
Moreover, the reinforcement learning framework is highly flexible and can incorporate
other reward functions such as T-gate count or circuit depth.
A key advantage of this reinforcement learning-based strategy is the scalability for
larger ZX-diagrams. Although the initial training phase can be
computationally demanding, the resulting agent can generalize its
learned strategies to new diagrams independent of their size.
Among the approaches surveyed, the combination of this reinforcement learning-based optimization
approach with the causal flow-preserving framework of \citet{holkerCausalFlowPreserving2024} represents one of
the most effective strategies to reduce the two-qubit gate counts in
ZX-based circuit optimization.

%Further adaptations to more complex heuristics
%could improve the learned strategies.

\section{Tree Search}
\subsection{Metric Agnostic}
\citet{fischbach2025exhaustivesearchquantumcircuit}~propose an tree search strategy combined with
pruning conditions to optimize ZX-diagrams. The
objective of the search is to find a sequence of rewriting rules that
optimizes a given metric.
Their method explores the state-space using depth-first search (DFS) and
iterative-deepening depth-first search (IDDFS) by applying
all possible rule combinations. Pruning conditions terminate branches
of the search tree, e.g., if circuit extraction is impossible.
Despite being a complete search for a rewriting rule sequence that
minimizes a given metric, the main drawback of their approach is the
computational cost.
A comparison of their IDDFS-based approach and the T-gate count
obtained by the full
reduce algorithm~\citet{duncanGraphtheoreticSimplificationQuantum2020} was performed.
The analysis demonstrated equivalent results in 89\% of the circuits
within a benchmark set of 100 structured circuits.
Nevertheless,  one strength of their approach is the
metric-agnosticism; it is not tied to the optimization of a single
metric such as the T-gate count. They demonstrated its flexibility by
targeting the edge count, a proxy for two-qubit gates, with the same strategy.
While the exhaustive nature of the search makes the approach
computationally inefficient for larger circuits, it allows for
optimal solutions for small sized circuits.
%Future work could focus
%on improving scalability through smarter heuristics, parallel
%exploration, or the optimization of small ZX-diagrams.

\citet{mattick2025optimizingquantumcircuitszx} propose a hybrid approach that combines reinforcement learning with a tree
search algorithm that uses the
full set of standard ZX-calculus
rules. Their
method slightly outperforms the stochastic and general flow preserving
techniques~\citet{staudacherReducing2QuBitGate2023} for random circuits. In
this framework, the graph neural network replaces the heuristics by
learning which and where a rewriting rule should be applied or
if the agent needs to stop. The tree search allows backtracking if
not beneficial transformations are encountered.
Similarly to~\citet{riuReinforcementLearningBased2023}, the
agent learns where to apply which rule in the ZX-diagram based on a
reward function that represents a metric. However, at each node of
the tree, the agent is allowed to perform only one diagram transformation.
The use of the complete set of ZX-calculus rewrite rules permits the
exploration of a larger state-space at the cost of post-processing to
ensure circuit extraction.
The key advantage of this approach is its generality. The framework
aims to learn optimal rewrite sequences for any chosen metric and is
not limited to minimizing two-qubit gate counts.

\section{Challenges}
\label{sec:challenges}

\paragraph{Comparability}
This survey lacks a comprehensive benchmark table that compares the various optimization strategies on the same set of  quantum circuits.
Most works are based on the PyZX~\cite{kissingerPyZXLargeScale2020} library.
As the result of PyZX's fast development cycle and substantial rewrites of the rewriting rule system and the graph data structure, most optimizers are only compatible with the specific PyZX version they were originally implemented in.

\paragraph{Scalability}
A major challenge in the optimization of ZX-diagrams lies in the
scalability of rule-based rewriting approaches, especially as
small real-world quantum circuits result in large ZX-diagrams. With
an increase in size of ZX-diagrams, the computational complexity
grows rapidly, limiting the practical application of existing
techniques to small-scale quantum circuits.

Current heuristic-based optimization strategies typically target a
single metric, such as reducing the number of non-Clifford spiders
(e.g.,~\citet{duncanGraphtheoreticSimplificationQuantum2020}, \citet{kissinger2020journal}, and \citet{beaudrap2020extended}) or  minimizing
the number of two-qubit gates (e.g.,~\citet{staudacherReducing2QuBitGate2023}, \citet{holkerCausalFlowPreserving2024}, and \citet{faganOptimisingCliffordCircuits2019}). However, while
heuristics improve
computational performance, focusing on a single objective fails to
capture the complex features of quantum circuits and how to balance
them. We identify the clear need for improved heuristics that can
balance multiple metrics to improve the computational performance of
existing strategies.

Recent works have explored reinforcement learning approaches for
diagram rewriting (\citet{nageleOptimizingZXDiagramsDeep2023}, \citet{riuReinforcementLearningBased2023}, \citet{mattick2025optimizingquantumcircuitszx}), demonstrating
promising results on small-scale
ZX-diagrams. Although reinforcement learning is still computationally expensive, the training
stage can be
seen as an upfront cost, as the learned strategies appear to be at
least partially
generalizable. Further enhancement of RL-based methods on small ZX-diagrams could allow better optimization results for large-scale circuits.
Moreover, the lack of interpretability and the theoretical background
as to why some rewriting sequences are more beneficial than others
poses additional
challenges. Incorporating explainable RL could form the basis for new
and computationally efficient heuristics that are applicable for large-scale
ZX-diagrams.

Another promising research direction is the introduction of
intermediate representations that can aggregate subdiagrams and
enable more efficient state-space exploration, similar to that of \citet{liu2024}.
The use of an intermediate representation allows template matching of
pre-optimized subcircuits.
\citet{chen2025quantumcircuitoptimizationbased}
dynamically group layers into subcircuits.
Future work should focus on the efficient partition and resynthesis
of quantum circuits for several reasons.
Dividing larger quantum circuits into smaller subcircuits is
beneficial because the resulting subdiagrams are smaller and can be
optimized efficiently.
As the different subdiagrams are independent of each other, each
instance can be solved in parallel.
This flexibility could lead the way for dynamic selection of the
optimization algorithm based on the characteristics of each subdiagram.
Furthermore, a subdiagram only needs to be optimized once and can be
substituted for further instances.
Another possible research direction is to allow for the two-dimensional
partition of quantum circuits for improved optimization results using
the intermediate representation of \citet{liu2024}.
Replacing the SA approach that changes the subcircuit partition of
\citet{chen2025quantumcircuitoptimizationbased} with an
tree search could further improve the results.

\paragraph{Architecture-Awareness} Quantum computing
architectures fundamentally differ from each other and offer
different advantages and
limitations that quantum circuit optimization needs to take into account.
There are four dominant quantum architectures currently considered by
ZX-based quantum circuit optimization:
(i) superconducting, (ii) trapped-ion, (iii) neutral-atom, and (iv)
photonic quantum architectures.

To create executable quantum circuits for superconducting quantum
computers spatial information (topology), qubit connectivity, noise,
and error correction.
Some work focuses on architecture-aware synthesis of phase
gadgets and polynomials (e.g., \citet{Meijer_van_de_Griend_2023}, \citet{gogiosoAnnealingOptimisationMixed2023}, and \citet{winderlRecursivelyPartitionedApproach2023}) that takes an
architecture's topology into account, others (e.g., \citet{kissinger2019cnotcircuitextractiontopologicallyconstrained} aim to restrict qubit connectivity).
However, there is no unified framework that combines the different methods.
Transpilation would be more efficient if ZX-diagram optimization
could target a specific architecture without the need for additional
optimization steps to consider decoherence, gate error, routing, and
error correction.
Phase gadgets are the native representation of multi-qubit gates in
ZX-calculus, therefore improving the architecture and topology aware
synthesis is a promising field for future research.

ZX-calculus is a natural candidate for photonic quantum computing because the architecture's measurement graph corresponds to a graph-like~(Definition~\ref{def:graph-like}) ZX-diagram.
This equivalence permits the modification of the measurement graph using rewriting rules.
As the measurement graph can be directly converted into hardware instructions, additional optimization steps are not required.
The initial work of \citet{zilk2022photonic} demonstrates the effectiveness of ZX-calculus to target photonic quantum computers.
Upcoming work could extend the approach of Zilke et al. to use different optimization algorithms for T-gate and Clifford gate elimination.
Especially the RL-approach of \citet{mattick2025optimizingquantumcircuitszx} seems to be a promising candidate for photonic architectures, as it balances the quality and exploration of the solution.
As photonic architectures do not require circuit extraction, the quality of the solution is not impacted by post-processing and circuit extraction.

%The initial work by \citet{staudacher2024neutralatom} proof the feasibility of ZX-calculus when targeting neutral-atom architectures.
%In their work, they adapt the circuit extraction algorithm of \citet{backensThereBackAgain2020} to efficiently synthesize the architecture's native multi-qubit gates.
%Forthcoming endeavors should include ZX-diagram optimization methods that take the reduction of global phase gates and the preference of multi-qubit gates before circuit extraction into account.

The standout features of trapped-ion quantum computers are the all-to-all qubit connectivity and the use of global gates.
\citet{villoria2025optimizationsynthesisquantumcircuits} modify the circuit extraction algorithm of \citet{backensThereBackAgain2020} to only extract vertices that take part in the same global gate.
Phase gadgets are the ZX-calculus equivalent of multi-qubit gates.
In the future, work could use the notion of phase gadgets during circuit extraction to improve global gate count and grouping.
Furthermore, there is no dedicated ZX-based optimization strategy
that targets trap-ion architectures outside of circuit extraction.

\paragraph{Circuit Extraction} A significant limitation of ZX-based quantum circuit optimization is the computational cost
of circuit extraction. In the general case, circuit extraction is \#P-hard~\cite{debeaudrapCircuitExtractionZXdiagrams2022,mitosek2024constructing}.
Although polynomial-time algorithms exist for ZX-diagrams that preserve the graph-theoretic conditions of general flow and causal flow, only a small subset of rules has been proven to preserve these flow properties.
In addition, verifying the presence of a flow is computationally expensive, with causal flow being less demanding than general flow.
Especially noteworthy are the extraction algorithms of \citet{duncanGraphtheoreticSimplificationQuantum2020} and its extension by \citet{backensThereBackAgain2020} that form the basis for various architecture-aware synthesis algorithms (e.g.,~\citet{kissinger2019cnotcircuitextractiontopologicallyconstrained}, \citet{villoria2025optimizationsynthesisquantumcircuits}, and \citet{staudacher2024neutralatom}).

We established that many approaches disregard intermediate ZX-diagrams without the presence of causal flow, general flow or Pauli flow, effectively ignoring large parts of the state-space that might contain the optimal solution. 
Future research should focus on methods that explore intermediate ZX-diagrams without preserving flow properties while  keeping the overhead introduced by circuit extraction at a minimum. This can be achieved by improving or avoiding circuit extraction as much as possible.

Based on \citet{quanz2024parallelextraction} subsequent
work should aim to improve parallel circuit extraction to speed up
other state-space exploration algorithms.

A promising research direction is to replace the Gaussian elimination algorithm of the biadjacency matrix during circuit extraction by a LP.
Similarly to \citet{villoria2025optimizationsynthesisquantumcircuits}, future
LP formulation could include architectural constraints.
Circuit extraction is an iterative process, and the biadjacency matrix
only captures the connectivity of the current frontier.
Therefore, the LP only encodes current information, and it is not
guaranteed that the optimal solution of the current LP results in the
best global solution.
Upcoming work should combine LP-based circuit extraction with a
backtracking algorithm that allows to prune LP solutions that result
in unfavorable quantum circuits.
Another way of providing context for LP-based circuit extraction, is
to provide information of the closest already extracted frontier
gates for each qubit.

It is important to recognize that the circuit properties strongly
depend on the circuit extraction algorithm itself.
Current circuit extraction algorithms replicate spider connectivity
by two-qubit gates, potentially increasing the circuit depth and
two-qubit gate count.
Improvements of circuit extraction algorithms, both in computational efficiency
and in circuit quality, should allow for better optimization results.
In particular, only the work of \citet{villoria2025optimizationsynthesisquantumcircuits} treats
circuit
extraction as a combinatorial problem.
Consequently, efficient
formulations beyond LP, constraints, and solvers for the circuit extraction
problem provide a vast field for future endeavors.

Many non-ad-hoc approaches require regular circuit extraction (e.g.,~\citet{fischbach2025exhaustivesearchquantumcircuit}, \citet{ewen2025genetic}, and \citet{riuReinforcementLearningBased2023}).
A critical challenge to avoid circuit extraction is that many
characteristics of quantum circuits, such as the two-qubit gate count
or circuit depth, are not native concepts of ZX-diagrams.
Investigating other approximations of quantum circuit metrics at the
ZX-diagram  level, such as Hadamard wires serving as a proxy for
two-qubit gates~\cite{staudacherReducing2QuBitGate2023}, might reduce
the amount of circuit extraction required.
A promising first step could be the construction of a surrogate model
for circuit extraction that can be quickly evaluated by different approaches.

\paragraph{Multi-Objective Optimization} The approaches presented in
this survey are designed to primarily target one metric.
Nevertheless, it does not suffice to optimize one metric alone to
capture the complexity of current quantum computing architectures.
Some works follow a lexicographic approach in which one metric after
another is improved.
A typical example is to first run the elimination of the T-gate
by~\citet{duncanGraphtheoreticSimplificationQuantum2020,kissinger2020journal} and then
optimize the two-qubit gate count on the resulting ZX-diagram
(e.g.~\citet{staudacherReducing2QuBitGate2023,holkerCausalFlowPreserving2024,faganOptimisingCliffordCircuits2019}).

The work of \citet{ewen2025genetic} suggests that the best
ZX-diagram results in a circuit close to its best known quantum
circuit counterpart for circuit depth and two-qubit gate count.
However, on the non-target metrics, the best circuits perform worse
than their ZX-based counterparts.

So far, there exist no deliberate multi-objective optimization approaches
applied to ZX-based quantum circuit optimization that aim to find a trade-off between
fundamentally different metrics.
Multi-objective optimization seems to be a promising candidate for
architecture-aware optimization where a tradeoff
between independent metrics and different architectural
specifications, such as qubit connectivity and spatial dimensions,
is required.
Bridging the gap towards multi-objective optimization would greatly simplify the transpilation pipeline, resulting into an integrated and potentially less computationally demanding framework.

\paragraph{ZX-Diagram Feature Encoding}
Despite many advances in ZX-diagram optimization and circuit extraction techniques, a fundamental question remains: what features of a ZX-diagram accurately describe the quality of the resulting quantum circuit? 
While some connections between the ZX-diagram and the quantum circuit are trivial, e.g., the number of $\frac{\pi}{4}$-phase spiders directly translates to the number of T-gates, other features are only an approximation or not translated at all.
The example we saw before was that the number of Hadamard edges and the overall spider connectivity serve as a proxy to estimate the
two-qubit gate count~\cite{staudacherReducing2QuBitGate2023}.

Further research on the mapping between ZX-diagram characteristics
and quantum circuit properties could enable the development of better optimization methods that explicitly account for architectural constraints.
Especially the inclusion cumulative gate error rates, qubit connectivity limitations, and coherence time would alleviate the need of a full transpilation pipeline. 
Such mappings could serve as a surrogate model that allows quicker heuristic or learning-based optimization without repeatedly requiring circuit extraction and transpilation to assess the quality of the solution.

Furthermore, we propose the systematic addition of characteristics that form a composite metric that aggregates information from features derived from the ZX-diagram, properties of the logical quantum circuit, and characteristics of the final transpiled and executable circuit. 
Investigating such composite metrics allows to dynamically  evaluate characteristics based on the current solution quality and computational cost.
Such approaches would allow identifying the relative importance of individual features in determining overall circuit quality. 
For example, it is unnecessary to take into account accumulated gate error and routing if the initial features already indicate an unfeasible solution.

This adaptive composite metric would allow excluding or adjust certain optimization steps based on the balance between the expected quality of the solution and the computational cost of fully evaluating that metric. 
By quantifying this trade-off, the composite metric could be added to many methods that require a cost function while also championing a full compilation workflow without being explicitly designed for it.

\section{Summary}
We provided a survey of quantum circuit optimization using ZX-calculus, with an emphasis on optimization techniques and target metrics.
The surveyed works demonstrate that ZX-calculus offers a powerful, compact, and universal framework that enables optimizations beyond traditional circuit-level techniques.
Furthermore, we identified that the adoption of ZX-based methods is impeded by scalability issues, a reliance on single-objective heuristics, and the computational cost of circuit extraction.

Several promising research directions emerge:
Future work must expand beyond single-metric optimization and adopt multi-objective optimization that jointly consider architecture-independent and architecture-aware metrics.
Additionally, there is a clear need to link ZX-diagram and quantum circuit characteristics, potentially through surrogate models or composite metrics that alleviate the computational cost of circuit extraction.
Furthermore, circuit extraction could be improved by leveraging combinatorial optimization with the potential inclusion of architectural constraints.
Finally, more approaches should take into account the underlying quantum computing architecture.
Especially trapped-ion, neutral-atom, and photonic devices are underrepresented.

Overall, ZX-calculus is positioned as an intermediate representation for circuit optimization, that is capable of bridging diagrammatic reasoning with architectural constraints of current and future quantum hardware. 
Improvements of heuristics, scalability, and explainable learning-based methods are necessary to design algorithms that handle efficiently the non-terminating nature of ZX-calculus.

In summary, ZX-calculus is a candidate for an integrated framework that allows architecture-independent and architecture-aware quantum circuit optimization for current and future quantum computing devices.

\bibliographystyle{plainnat}
\bibliography{zx_survey}

\end{document}